\RequirePackage{commath, amsmath, amssymb, amsfonts}
\documentclass[10pt, a4paper]{iopart}

\usepackage{xcolor}
\usepackage{physics}
\usepackage{graphicx}
\usepackage{makecell}
\graphicspath{{fig/}}

\newcommand{\GW}{G$_0$W$_0$ }

\begin{document}

\title{Recent Progress of the Computational 2D Materials Database (C2DB)}
%New materials, properties, and features of the Computational 2D Materials Database (C2DB)}

\author{Morten Niklas Gjerding$^1$, Alireza Taghizadeh$^2,^1$, Asbjørn Rasmussen$^1$, Sajid Ali$^1$, Fabian Bertoldo$^1$, Thorsten Deilmann$^3$, Urko Petralanda Holguin$^1$, Nikolaj Rørbæk Knøsgaard$^1$, Mads Kruse$^1$, Simone Manti$^1$, Thomas Garm Pedersen$^2$, Thorbjørn Skovhus$^1$, Mark Kamper Svendsen$^1$, Jens Jørgen Mortensen$^1$, Thomas Olsen$^1$ and Kristian Sommer Thygesen$^1$}
\address{$^1$ Computational Atomic-scale Materials Design (CAMD), Department of Physics, Technical University of Denmark, 2800 Kgs. Lyngby Denmark}
\address{$^2$ Department of Materials and Production, Aalborg University, 9220 Aalborg {\O}st, Denmark}
\address{$^3$ Institut für Festk\"orpertheorie, Westfälische Wilhelms-Universit\"at M\"unster, 48149 M\"unster, Germany}
\ead{thygesen@fysik.dtu.dk}
\vspace{10pt}

\begin{abstract}
The C2DB is a highly curated open database organizing a wealth of computed properties for more than 4000 atomically thin two-dimensional (2D) materials. Here we report on new materials and properties that were added to the database since its first release in 2018. The set of new materials comprise several hundred monolayers exfoliated from experimentally known layered bulk materials, (homo)bilayers in various stacking configurations, native point defects in semiconducting monolayers, and chalcogen/halogen Janus monolayers. The new properties include exfoliation energies, Bader charges, spontaneous polarisations, Born charges, infrared polarisabilities, piezoelectric tensors,  band topology invariants, exchange couplings, Raman- and second harmonic generation spectra. We also describe refinements of the employed material classification schemes, upgrades of the computational methodologies used for property evaluations, as well as significant enhancements of the data documentation and provenance. Finally, we explore the performance of Gaussian process-based regression for efficient prediction of mechanical and electronic materials properties. The combination of open access, detailed documentation, and extremely rich materials property data sets make the C2DB a unique resource that will advance the science of atomically thin materials.
\end{abstract}

\submitto{\TDM}
\maketitle
\ioptwocol

\section{Introduction}
The discovery of new materials, or new properties of known materials, to meet a specific industrial or scientific requirement, is an exciting intellectual challenge of the utmost importance for our environment and economy. For example, the successful transition to a society based on sustainable energy sources and the realisation of quantum technologies (e.g. quantum computers and quantum communication) depend critically on new materials with novel functionalities. First-principles quantum mechanical calculations, e.g. based on density functional theory (DFT)\cite{kohn1965self}, can predict the properties of materials with high accuracy even before they are made in the lab. They provide insight into mechanisms at the most fundamental (atomic and electronic) level and can pinpoint and calculate key properties that determine the performance of the material at the macroscopic level. Powered by high-performance computers, atomistic quantum calculations in combination with data science approaches, have the potential to revolutionize the way we discover and develop new materials.   

Atomically thin, two-dimensional (2D) crystals represent a fascinating class of materials with exciting perspectives for both fundamental science and technology\cite{schwierz2010graphene,novoselov20162d,ferrari2015science,bhimanapati2015recent}. The family of 2D materials has been growing steadily over the past decade and counts about a hundred materials that have been realised in single- or few-layer form\cite{haastrup2018computational,shivayogimath2019universal,zhou2018library,anasori20172d,dou2015atomically}. While some of these materials, including graphene, hexagonal boron-nitride (hBN), and transition metal dichalcogenides (TMDs), have been extensively studied, the majority have only been scarcely characterized and remain poorly understood. Computational studies indicate that around 1000 already known layered crystals have sufficiently weak interlayer bonding to allow the individual layers to be mechanically exfoliated\cite{mounet2018two,ashton2017topology}. Supposedly, even more 2D materials could be realized beyond this set of already known crystals. Adding to this the possibility of stacking individual 2D layers (of the same or different kinds) into ultrathin van der Waals (vdW) crystals\cite{geim2013van}, and tuning the properties of such structures by varying the relative twist angle between adjacent layers\cite{cao2018unconventional,bistritzer2011moire} or intercalating atoms into the vdW gap\cite{zhao2020engineering,wan2016tuning}, it is clear that the prospects of tailor made 2D materials are simply immense. To support experimental efforts and navigate the vast 2D materials space, first-principles calculations play a pivotal role. In particular, FAIR\footnote{FAIR data are data which meet principles of findability, accessibility, interoperability, and reusability}\cite{wilkinson2016fair} databases populated by high-throughput calculations can provide a convenient overview of known materials and point to new promising materials with desired (predicted) properties. Such databases are also a fundamental requirement for the successful introduction and deployment of artificial intelligence in materials science.    

Many of the unique properties exhibited by 2D materials have their origin in quantum confinement and reduced dielectric screening. These effects tend to enhance many-body interactions and lead to profoundly new phenomena such as strongly bound excitons\cite{wirtz2006excitons,cudazzo2011dielectric,klots2014probing} with nonhydrogenic Rydberg series\cite{chernikov2014exciton,olsen2016simple,riis2020anomalous}, phonons and plasmons with anomalous dispersion relations\cite{felipe2020universal,sohier2017breakdown}, large dielectric band structure renormalizations\cite{ugeda2014giant,winther2017band}, unconventional Mott insulating and superconducting phases\cite{cao2018unconventional,bistritzer2011moire}, and high-temperature exciton condensates\cite{wang2019evidence}. Recently, it has become clear that long range magnetic order can persist\cite{gong2017discovery,huang2017layer} and (in-plane) ferroelectricity even be enhanced\cite{chang2016discovery}, in the single layer limit. In addition, first-principles studies of 2D crystals have revealed rich and abundant topological phases\cite{olsen2019discovering,marrazzo2019relative}. The peculiar physics ruling the world of 2D materials entails that many of the conventional theories and concepts developed for bulk crystals break down or require special treatments when applied to 2D materials\cite{thygesen2017calculating,sohier2017breakdown,Torelli2019}. This means that computational studies must be performed with extra care, which in turn calls for well-organized and well-documented 2D property data sets that can form the basis for the development, benchmarking, and consolidation of physical theories and numerical implementations.   

The Computational 2D Materials Database (C2DB)\cite{haastrup2018computational,rasmussen2015computational} is a highly curated and fully open database containing elementary physical properties of around 4000 two-dimensional (2D) monolayer crystals. The data has been generated by automatic high-throughput calculations at the level of density functional theory (DFT) and many-body perturbation theory as implemented in the GPAW electronic structure code. The computational workflow is constructed using the Atomic Simulation Recipes (ASR) -- a recently developed Python framework for high-throughput materials modeling building on the Atomic Simulation Environment (ASE) -- and managed/executed using the MyQueue task scheduler\cite{Mortensen2020}. 

The C2DB differentiates itself from existing computational databases of bulk\cite{saal2013materials,jain2013commentary,curtarolo2012aflow} and low-dimensional\cite{ashton2017topology,mounet2018two,zhou20192dmatpedia} materials, by the large number of physical properties available, see Table \ref{tab:properties}. The use of beyond-DFT theories for excited state properties (GW band structures and BSE absorption for selected materials) and Berry-phase techniques for band topology and polarization quantities (spontaneous polarization, Born charges, piezoelectric tensors), are other unique features of the database. 

The C2DB can be downloaded in its entirety or browsed and searched online. As a new feature, all data entries presented on the website are accompanied by a clickable help icon that presents a scientific documentation ("what does this piece of data describe?") and technical documentation ("how was this piece of data computed?"). This development enhances the usability of the database and improves the reproducibility and provenance of the data contained in C2DB. As another novelty it is possible to download all property data pertaining to a specific material or a specific type of property, e.g. the band gap, for all materials thus significantly improving data accessibility.        
 
In this paper, we report on the significant C2DB developments that have taken place during the past two years. These developments can be roughly divided into four categories: (i) General updates of the workflow used to select, classify, and stability assess the materials. (ii) Computational improvements for properties already described in the 2018 paper. (iii) New properties. (vi) New materials. The developments, described in four separate sections, cover both original work and review of previously published work. In addition, we have included some outlook discussions of ongoing work. In the last section we illustrate an application of statistical learning to predict properties directly from the atomic structure.

\section{Selection, classification, and stability}\label{sec:select}
\begin{figure*}[t]
    \centering
    \includegraphics[width=\textwidth]{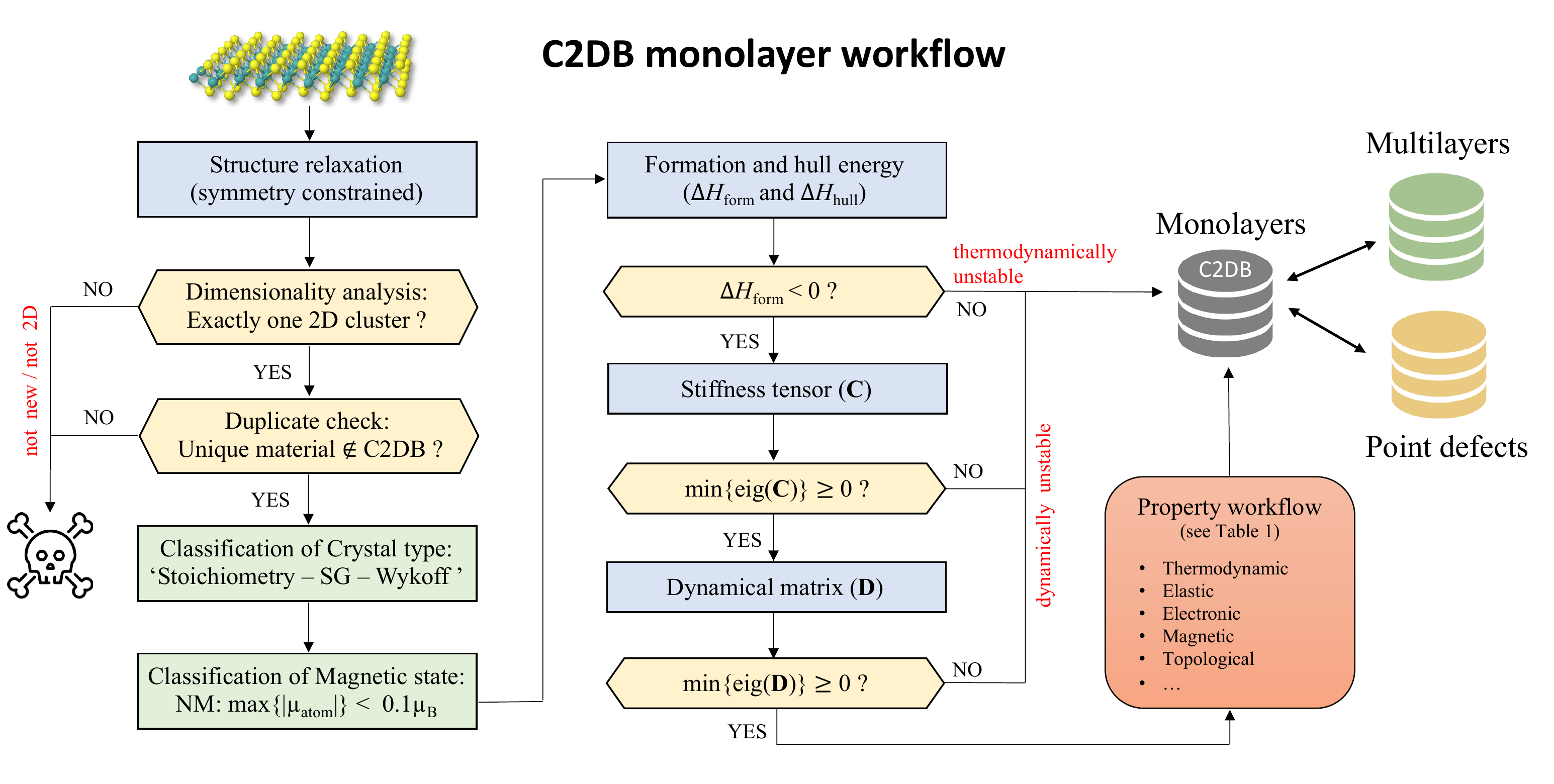}
    \caption{
      The workflow behind the C2DB. After the structural relaxation, the dimensionality of the material is checked and it is verified that the material is not already present in the database. Next, the material is classified according to its chemical composition, crystal structure, and magnetic state. Finally, the thermodynamic- and dynamic stability is assessed from the energy above the convex hull and the sign of the minimum eigenvalues of the dynamical matrix and stiffness tensor. Unstable materials are stored in the database; stable materials are subject to the property workflow. The C2DB monolayer database is interlinked with databases containing structures and properties of multilayer stacks and point defects in monolayers from the C2DB.  
    }
    \label{fig:workflow}
\end{figure*}
Figure \ref{fig:workflow} illustrates the workflow behind the C2DB. In this section we describe the first part of the workflow until the property calculations (red box), focusing on aspects related to selection criteria, classification, and stability assessment, that have been changed or updated since the 2018 paper.

\begin{table*}[!ht]
    \begin{tabular}{l|l|l|l}
        Property & Method & Criteria & Count  \\ \hline\hline
        Bader charges & PBE &  & 3809 \\ \hline
        Energy above convex hull & PBE &  & 4044 \\ \hline
        Heat of formation & PBE &  & 4044 \\ \hline
        Orbital projected band structure & PBE &   & 2487 \\ \hline
        Out-of-plane dipole & PBE &  & 4044 \\ \hline
        Phonons ($\Gamma$ and BZ corners) & PBE &  & 3865 \\ \hline
        Projected density of states & PBE &  & 3332 \\ \hline
        Stiffness tensor & PBE &  & 3968 \\ \hline
        Exchange couplings & PBE & Magnetic & 538 \\ \hline
        Infrared polarisability & PBE & $E^\mathrm{PBE}_{\mathrm{gap}}>0$ & 784 \\ \hline
        Second harmonic generation & PBE & \makecell[lt]{$E^\mathrm{PBE}_{\mathrm{gap}}>0$, non-magnetic, \\ non-centrosymmetric  } & 375 \\ \hline
        Electronic band structure PBE & PBE* &  & 3496 \\ \hline
        Magnetic anisotropies & PBE* & Magnetic & 823 \\ \hline
        Deformation potentials & PBE* & $E^\mathrm{PBE}_{\mathrm{gap}}>0$ & 830 \\ \hline
        Effective masses & PBE* & $E^\mathrm{PBE}_{\mathrm{gap}}>0$ & 1272 \\ \hline
        Fermi surface & PBE* & $E^\mathrm{PBE}_{\mathrm{gap}}=0$ & 2505 \\ \hline
        Plasma frequency & PBE* & $E^\mathrm{PBE}_{\mathrm{gap}}=0$ & 3144 \\ \hline
        Work function & PBE* & $E^\mathrm{PBE}_{\mathrm{gap}}=0$ & 4044 \\ \hline
        Optical polarisability & RPA@PBE &  & 3127 \\ \hline
        Electronic band structure & HSE06@PBE* &  & 3155 \\ \hline
        Electronic band structure & G$_0$W$_0$@PBE* & $E^\mathrm{PBE}_{\mathrm{gap}}>0$, $N_{\mathrm{atoms}}<5$ & 357 \\ \hline
        Born charges & PBE, Berry phase & $E^\mathrm{PBE}_{\mathrm{gap}}>0$ & 639 \\ \hline
        Raman spectrum & PBE, LCAO basis set & Non-magnetic, dyn. stable & 708 \\ \hline
        Piezoelectric tensor & PBE, Berry phase & $E^\mathrm{PBE}_{\mathrm{gap}}$, non-centrosym. & 353 \\ \hline
        Optical absorbance & BSE@G$_0$W$_0$* & $E^\mathrm{PBE}_{\mathrm{gap}}>0$,  $N_{\mathrm{atoms}}<5$ & 378 \\ \hline
        Spontaneous polarisation & PBE, Berry phase & \makecell[lt]{$E^\mathrm{PBE}_{\mathrm{gap}}>0$, nearly centrosym.\\polar space group} & 151 \\ \hline
        Topological invariants & PBE*, Berry phase & $0<E^\mathrm{PBE}_{\mathrm{gap}}<0.3$ eV & 242
    \end{tabular}
    \caption{Properties calculated by the C2DB monolayer workflow. The computational method and the criteria used to decide whether the property should be evaluation for a given material is also shown. A '*' indicates that spin-orbit coupling (SOC) is included. All calculations are performed with the GPAW code using a plane wave basis except for the Raman calculations, which employ a double-zeta polarized (DZP) basis of numerical atomic orbitals. }
    \label{tab:properties}
\end{table*}

\subsection{Structure relaxation}\label{sec:relax}
Given a prospect 2D material, the first step is to carry out a structure optimization. This calculation is performed with spin polarization and with the symmetries of the original structure enforced. The latter is done to keep the highest level of control over the resulting structure by avoiding ``uncontrolled'' symmetry breaking distortions. The prize to pay is a higher risk of generating dynamically unstable structures.    

\subsection{Selection: Dimensionality analysis}\label{sec:dimensionality}
\begin{figure}[t]
    \centering
    \includegraphics[width=\linewidth]{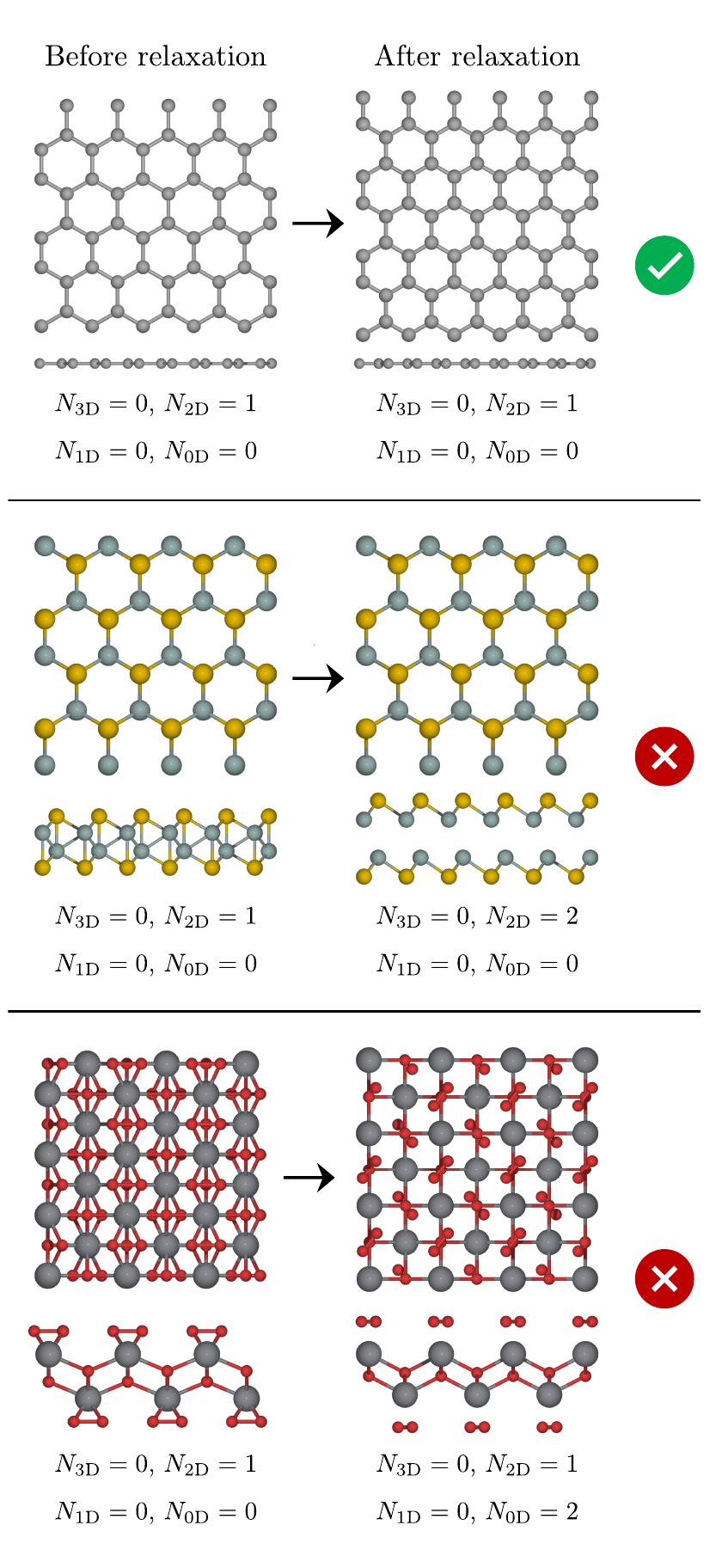}
    \caption{Three example structures from C2DB (top: graphene, middle: Ge$_2$Se$_2$, bottom: Pb$_2$O$_6$) with their respective cluster dimensionalities cluster before (left) and after (right) relaxation. The number $N_{x\mathrm{D}}$ denotes the number of clusters of dimensionality $x$.}
    \label{fig:dimensionality}
\end{figure}

A dimensionality analysis\cite{Larsen2019} is performed to identify and filter out materials that have disintegrated into non-2D structures during relaxation. Covalently bonded clusters are identified through an analysis of the connectivity of the structures where two atoms are considered to belong to the same cluster if their distance is less than some scaling of the sum of their covalent radii, i.e. $d<k(r_i^\mathrm{cov} + r_j^\mathrm{cov})$, where $i$ and $j$ are atomic indices. A scaling factor of $k=1.35$ was determined empirically. Only structures that consist of a single 2D cluster after relaxation are further processed. Figure \ref{fig:dimensionality} shows three examples (graphene, Ge$_2$Se$_2$, and Pb$_2$O$_6$) of structures and their cluster dimensionalities before and after relaxation. All structures initially consist of a single 2D cluster, but upon relaxation Ge$_2$Se$_2$ and Pb$_2$O$_6$ disintegrate into two 2D clusters as well as one 2D and two 0D clusters, respectively. On the other hand, the relaxation of graphene decreases the in-plane lattice constant but does not affect the dimensionality. According to the criterion defined above only graphene will enter the database.

\subsection{Selection: Ranking similar structures}\label{sec:similar}
 Maintaining a high-throughput database inevitably requires a strategy for comparing similar structures and ranking them according to their relevance. In particular, this is necessary in order to identify different representatives of the same material e.g. resulting from independent relaxations, and thereby avoid duplicate entries and redundant computations. The C2DB strategy to this end involves a combination of structure clustering and Pareto analysis. 
 
 First, a single-linkage  clustering  algorithm is used to group materials with identical reduced chemical formula and "similar" atomic configurations. To quantify configuration similarity a slightly modified version of PyMatGen's\cite{ong2013python} distance metric is employed where the cell volume normalization is removed to make it applicable to 2D materials surrounded by vacuum. Roughly speaking, the metric measures the maximum distance an atom must be moved (in units of \AA) in order to match the two atomic configurations. Two atomic configurations belong to the same cluster if their distance is below an empirically determined threshold of 0.3 \AA.

\begin{figure*}[t]
    \centering
    \includegraphics{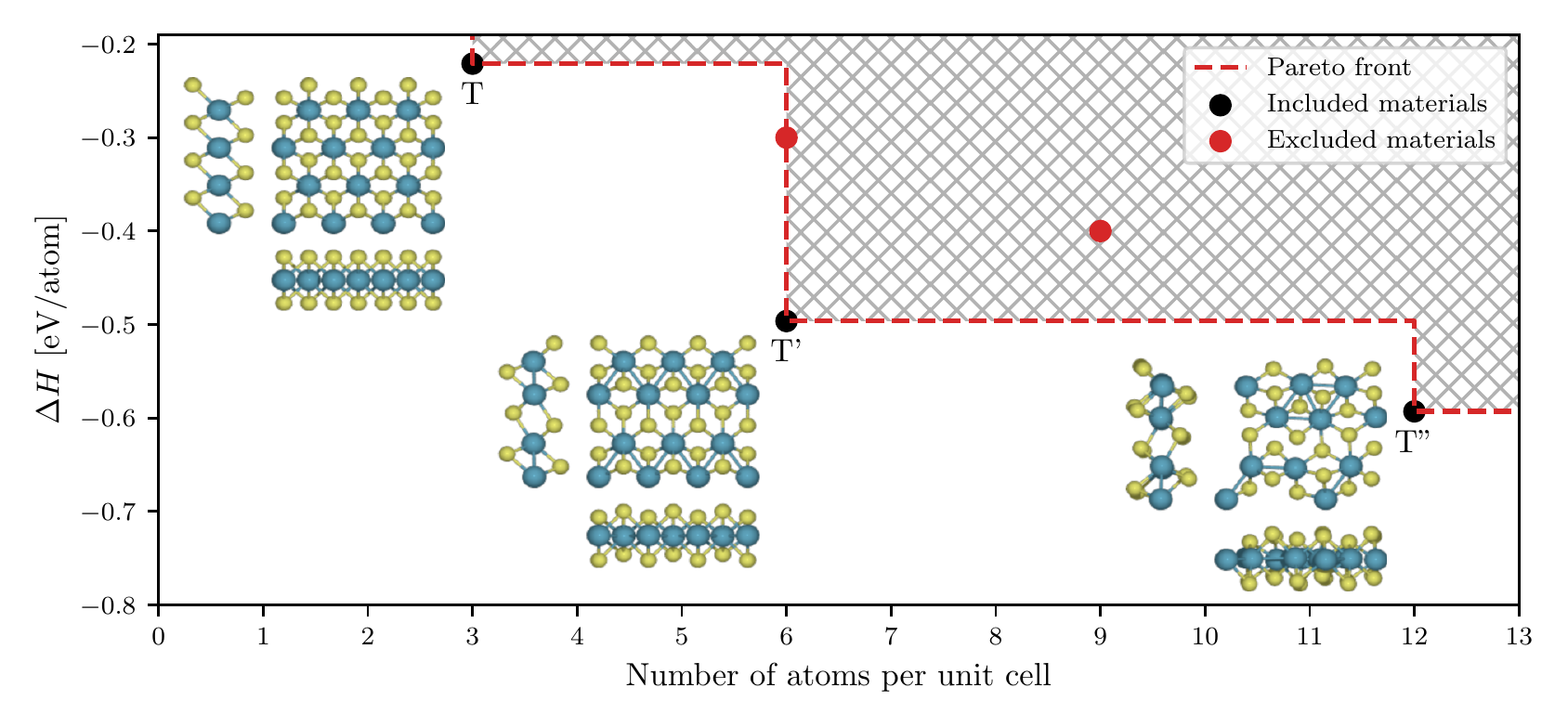}
    \caption{Illustration of the Pareto analysis used to filter out duplicates or irrelevant structures from the C2DB. All points represent materials with the same reduced chemical formula (in this case ReS$_2$) that belong to the same cluster defined by the structure metric. Only structures lying on the ($N,\Delta H$)-Pareto front are retained (black circles) while other materials are excluded (red circles). The philosophy behind the algorithm is to keep less stable materials if they contain fewer atoms per unit cell than more stable materials and thus represent structures of higher symmetry. 
    }
    \label{fig:duplicates-pareto}
\end{figure*}

At this point, the simplest strategy would be to remove all but the most stable compound within a cluster. However, this procedure would remove many high symmetry crystals for which a more stable distorted version exists. For example, the well known T-phase of MoS$_2$ would be removed in favor of the more stable T'-phase. This is undesired as high-symmetry structures, even if dynamically unstable at $T=0$, may provide useful information and might in fact become stabilized at higher temperatures\cite{patrick2015anharmonic}. Therefore, the general strategy adopted for the C2DB, is to keep a material that is less stable than another material of the same cluster if it has fewer atoms in its primitive unit cell (and thus typically higher symmetry). Precisely, materials within a given cluster are kept only if they represent a defining point of the ($N$, $\Delta H$)-Pareto front, where $N$ is the number of atoms in the unit cell and $\Delta H$ is the heat of formation. A graphical illustration of the Pareto analysis is shown in Figure \ref{fig:duplicates-pareto} for the case of ReS$_2$.

\subsection{Classification: Crystal structure}
The original C2DB employed a \emph{crystal prototype} classification scheme where specific materials were promoted to prototypes and used to label groups of materials with the same or very similar crystal structure. This approach was found to be difficult to maintain (as well as being non-transparent). Instead, materials are now classified according to their \textit{crystal type} defined by the reduced stoichiometry, space group number, and the alphabetically sorted labels of the occupied Wyckoff positions. As an example, MoS$_2$ in the H-phase has the crystal type: AB2-187-bi.

% Note: Prototype means an individual that exhibits the essential features of a later type / a standard or typical example. 

\subsection{Classification: Magnetic state}\label{sec:magnetic}
In the new version of the C2DB, materials are classified according to their magnetic state as either \emph{non-magnetic} or \emph{magnetic}. A material is considered magnetic if any atom has a local magnetic moments greater than 0.1 $\mu_\mathrm{B}$. 

In the original C2DB, the \emph{magnetic} category was further subdivided into ferromagnetic (FM) and anti-ferromagnetic (AFM). But since the simplest anti-ferromagnetically ordered state typically does not represent the true ground state, all material entries with an AFM state have been removed from the C2DB and replaced by the material in its FM state. Although the latter is less stable,  it represents a more well defined state of the material. Crucially, the nearest neighbor exchange couplings for all magnetic materials have been included in the C2DB (see Sec. \ref{sec:exchange}). This enables a more detailed and realistic description of the magnetic order via the Heisenberg model. In particular, the FM state of a material is not expected to represent the true magnetic ground if the exchange coupling $J<0$.

% First spin polarized. Switch to non-spin polarized if XXX. We keep only the most stable phase (different from original where both FM and NM were kept if close in energy). We do not consider AFM states, since 1) there can be many different AFM configurations and it is cumbersome to check all. 2) in reality perfect AFM order is rare. Instead we calculate J and check the sign (see section ?) 

\subsection{Stability: Thermodynamic}
\begin{figure}
    \centering
    \includegraphics{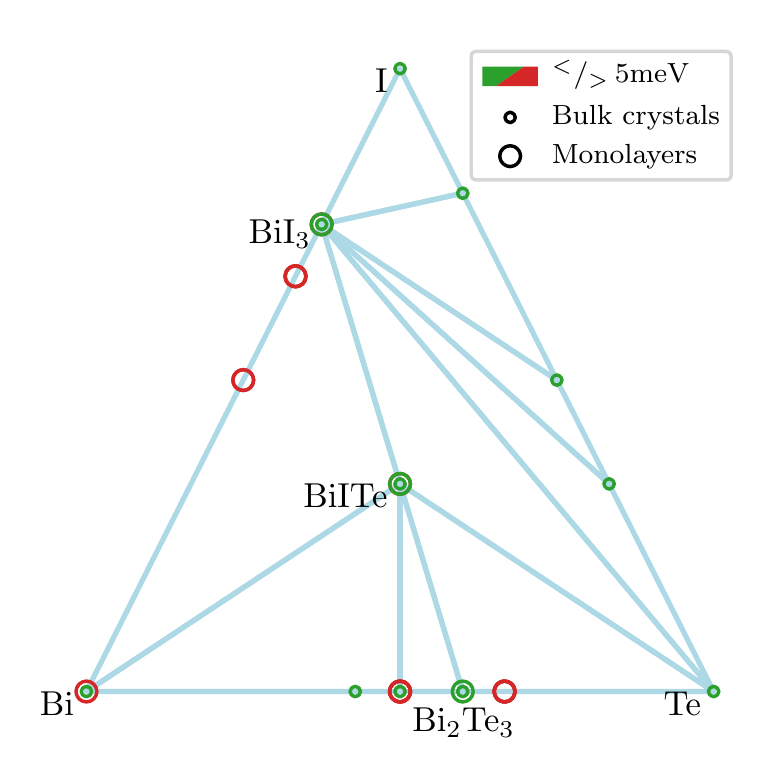}
    \caption{Convex hull diagram for (Bi,I,Te)-compounds. Green (red) coloring indicate materials that have a convex hull energy of less than (greater than) 5 meV. The monolayers BiI$_3$,  Bi$_2$Te$_3$ and BiITe lie on the convex hull. The monolayers are degenerate with their layered bulk parent because the vdW interactions are not captured by the PBE xc-functional.}
    \label{fig:convex-hull}
\end{figure}

The heat of formation, $\Delta H$, of a compound is defined as its energy per atom relative to its constituent elements in their standard states.\cite{kirklin2015open} The thermodynamic stability of a compound is evaluated in terms of its energy above the \emph{convex hull}, $\Delta H_{\mathrm{hull}}$, which gives the energy of the material relative to other competing phases of the same chemical composition, including mixed phases\cite{haastrup2018computational}, see Fig. \ref{fig:convex-hull} for an example. Clearly, $\Delta H_{\mathrm{hull}}$, depends on the pool of reference phases, which in turn defines the convex hull. The original C2DB employed a pool of reference phases comprised by 2807 elemental and binary bulk crystals from the convex hull of the Open Quantum Materials Database (OQMD)\cite{kirklin2015open}. In the new version, this set has been extended by approximately 6783 ternary bulk compounds from the convex hull of OQMD, making a total of 9590 stable bulk reference compounds.  

As a simple indicator for the thermodynamic stability of a material, the C2DB employs three labels (low, medium, high) as defined in Table \ref{tab:stability}. 

\begin{table}
    \begin{tabular}{l|l}
    \makecell[l]{Thermodynamic\\ stability indicator} & Criterion \\ \hline \hline
    LOW & $\Delta H>0$ \\
    MEDIUM & $\Delta H<0$ and $\Delta H_{\mathrm{hull}}>0.2$ eV\\
    HIGH &  $\Delta H<0$ and $\Delta H_{\mathrm{hull}}<0.2$ eV\\ \hline
    
    \end{tabular}
    \caption{Thermodynamic stability indicator assigned to all materials in the C2DB.}
    \label{tab:stability}
\end{table}

It should be emphasized that the energies of both monolayers and bulk reference crystals are calculated with the PBE xc-functional. This implies that some inaccuracies must be expected, in particular for materials with strongly localized $d$-electrons, e.g. certain transition metal oxides, and materials for which dispersive interactions are important, e.g. layered van der Waals crystals. The latter implies that the energy of a monolayer and its layered bulk parent (if such exists in the pool of references) will have the same energy. For further details and discussions see Ref. \cite{haastrup2018computational}.

\subsection{Stability: Dynamical}\label{sec:dynamical}
Dynamically stable materials are situated at a local minimum of the potential energy surface and are thus stable to small structural perturbations. Structures resulting from DFT relaxations can end up in saddle point configurations because of imposed symmetry constraints or an insufficient number of atoms in the unit cell. 

In C2DB, the dynamical stability is assessed from the signs of the minimum eigenvalues of (i) the stiffness tensor (see Sec. \ref{sec:stiffness}) and (ii) the $\Gamma$-point Hessian matrix for a supercell containing $2\times 2$ repetitions of the unit cell (the structure is not relaxed in the $2\times 2$ supercell). If one of these minimal eigenvalues is negative the material is classified as dynamically unstable. This indicates that the energy can be reduced by displacing an atom and/or deforming the unit cell, respectively. The use of two categories for dynamical stability, i.e. stable/unstable, differs from the original version of the C2DB where an intermediate category was used for materials with negative but numerically small minimal eigenvalue of either the Hessian or stiffness tensors.

\section{Improved property methodology}\label{sec:improvedprop}
The new version of the C2DB has been generated using a significantly extended and improved workflow for property evaluations. This section focuses on improvements relating to properties that were already present in the original version of the C2DB while new properties are discussed in the next section. 

\subsection{Stiffness tensor}\label{sec:stiffness}
The stiffness tensor, $C$, is a rank-4 tensor that relates the stress of a
material to the applied strain. In Mandel notation (a variant of Voigt notation) $C$ is expressed as an $N\times N$ matrix relating the $N$ independent components of the stress and strain tensors. For a 2D material $N=3$ and the tensor takes the form
\begin{align}
    \mathbf{C} = 
    \begin{bmatrix}
    C_{xxxx} & C_{xxyy} & \sqrt{2}C_{xxxy}\\
    C_{xxyy} & C_{yyyy} & \sqrt{2}C_{yyxy}\\
    \sqrt{2}C_{xxxy} & \sqrt{2}C_{yyxy} & 2C_{xyxy}
    \end{bmatrix}
\end{align}
where the indices on the matrix elements refer to the rank-4 tensor. The factors multiplying the tensor elements account for their multiplicities in the full rank-4 tensor. In the C2DB workflow, $C$ is calculated as a finite difference of the stress under an applied strain with full relaxation of atomic coordinates. A negative eigenvalue of $C$ signals a dynamical instability, see Sec. \ref{sec:dynamical}

In the first version of the C2DB only the diagonal elements of the stiffness tensor were calculated. The new version also determines the shear components such that the full $3\times 3$ stiffness tensor is now available. This improvement also leads to a more accurate assessment of dynamical stability\cite{Madziarz2019}.

\subsection{Effective masses with parabolicity estimates}
For all materials with a finite band gap the effective masses of electrons and holes are calculated for bands within 100 meV of the conduction band minimum (CBM) and valence band maximum (VBM), respectively. The Hessian matrices at the band extrema (BE) are determined by fitting a second order polynomium to the PBE band structure including SOC, and the effective masses are obtained by subsequent diagonalization of the Hessian. The main fitting-procedure is unaltered from the first version of C2DB, but two important improvements have been made.

The first improvement consists in an additional $k$-mesh refinement step for better localization of the BE in the Brillouin zone. After the location of the BE have been estimated based on a uniformly sampled band structure with $k$-point density of 12 Å, another one-shot calculation is perform with a denser $k$-mesh around the estimated BE positions. This ensures a more accurate and robust determination of the location of the BE, which can be important in cases with a small but still significant spin-orbit splitting or when the band is very flat or non-quadratic around the BE. The second refinement step is the same as in the first version of C2DB, i.e. the band energies are calculated on a highly dense $k$-mesh in a small disc around the BE, and the Hessian is obtained by fitting the band energies in the range up to 1 meV from the band minimum/maximum.

The second improvement is the calculation of the mean absolute relative error (MARE) of the polynomial fit in a 25 meV range from the band minimum/maximum. The value of 25 meV corresponds to the thermal energy at room temperature and is thus the relevant energy scale for many applications. The MARE provides a useful measure of the parabolicity of the energy bands and thus the validity of the effective mass approximation over this energy scale.   

\begin{figure*}[t]
    \centering
    \includegraphics{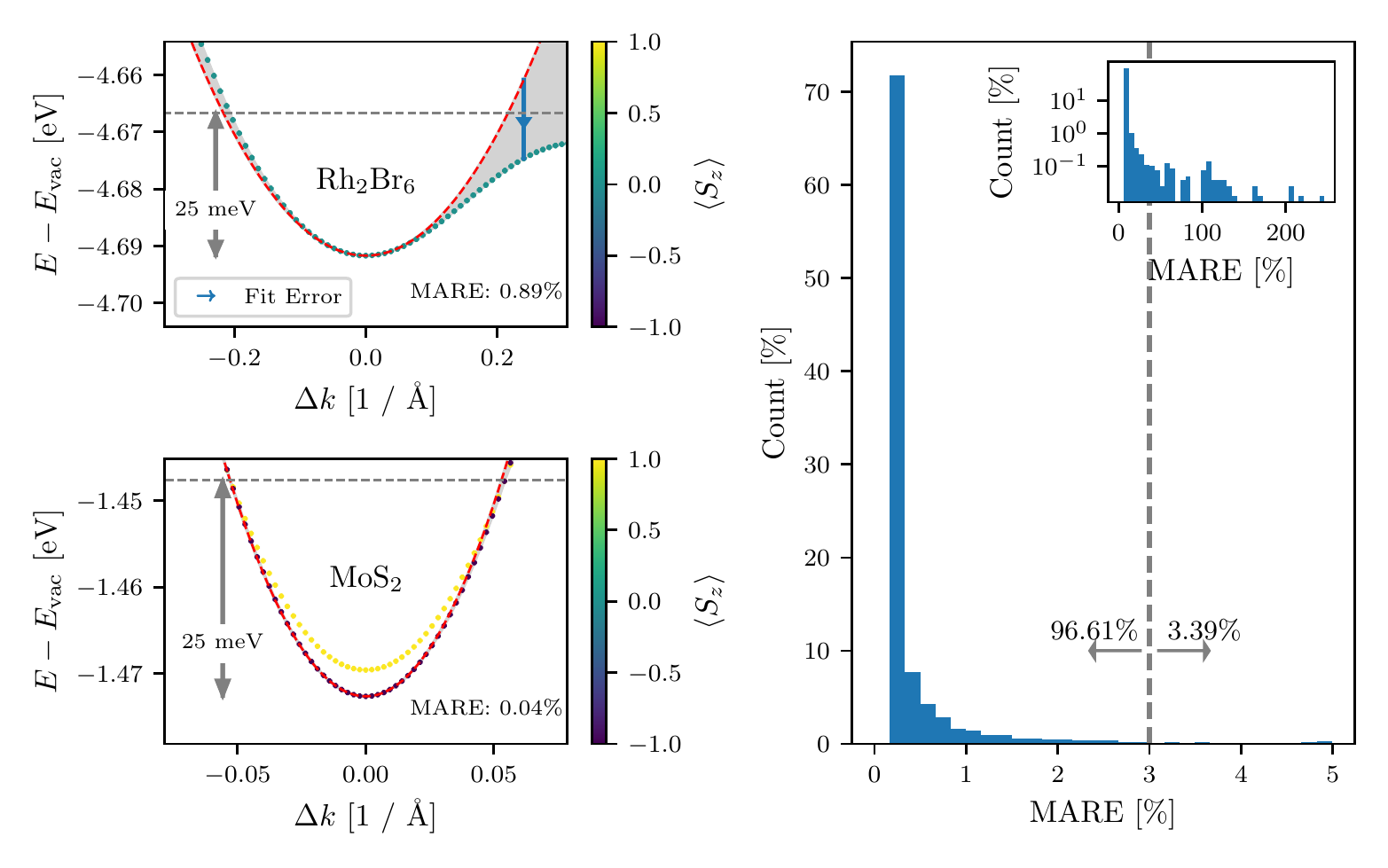}
    \caption{Left: The PBE band structures of Rh$_2$Br$_6$ and MoS$_2$ (colored dots) in regions around the conduction band minimum. The dashed red line shows the fit made to estimate the effective masses of the lowest conduction band. The shaded grey region highlights the error between the fit and the true band structure. The mean absolute relative error (MARE) is calculated for energies within 25 meV of the band minimum. For MoS$_2$ the fit is essentially ontop of the band energies. Right: The distribution of the MARE of all effective mass fits in the C2DB. The inset shows the full distribution on a log scale. As mentioned in the main text, very large MAREs indicate that the band minimum/maximum was incorrectly identified by the algorithm and/or that the band is very flat.}
    \label{fig:emasses}
\end{figure*}

Figure \ref{fig:emasses} shows two examples of band structures with the effective mass fits and corresponding fit errors indicated. Additionally, the distribution of MARE for all the effective mass fits in the C2DB are presented. Most materials have an insignificant MARE, but a few materials have very large errors. Materials with a MARE above a few tens of percentages fall into two classes. For some materials the algorithm does not correctly find the position of the BE. An example is Ti$_2$S$_2$ in the space group C2/m. For others, the fit and BE location are both correct, but the band flattens away from the BE which leads to a large MARE. An example of this latter class is Cl$_2$Tl$_2$ in the space group P-1. In general a small MARE indicates a parabolic band while materials with large MARE should be handled on a case-by-case basis.

\subsection{Orbital projected band structure}

\begin{figure*}[t]
    \centering
    \includegraphics{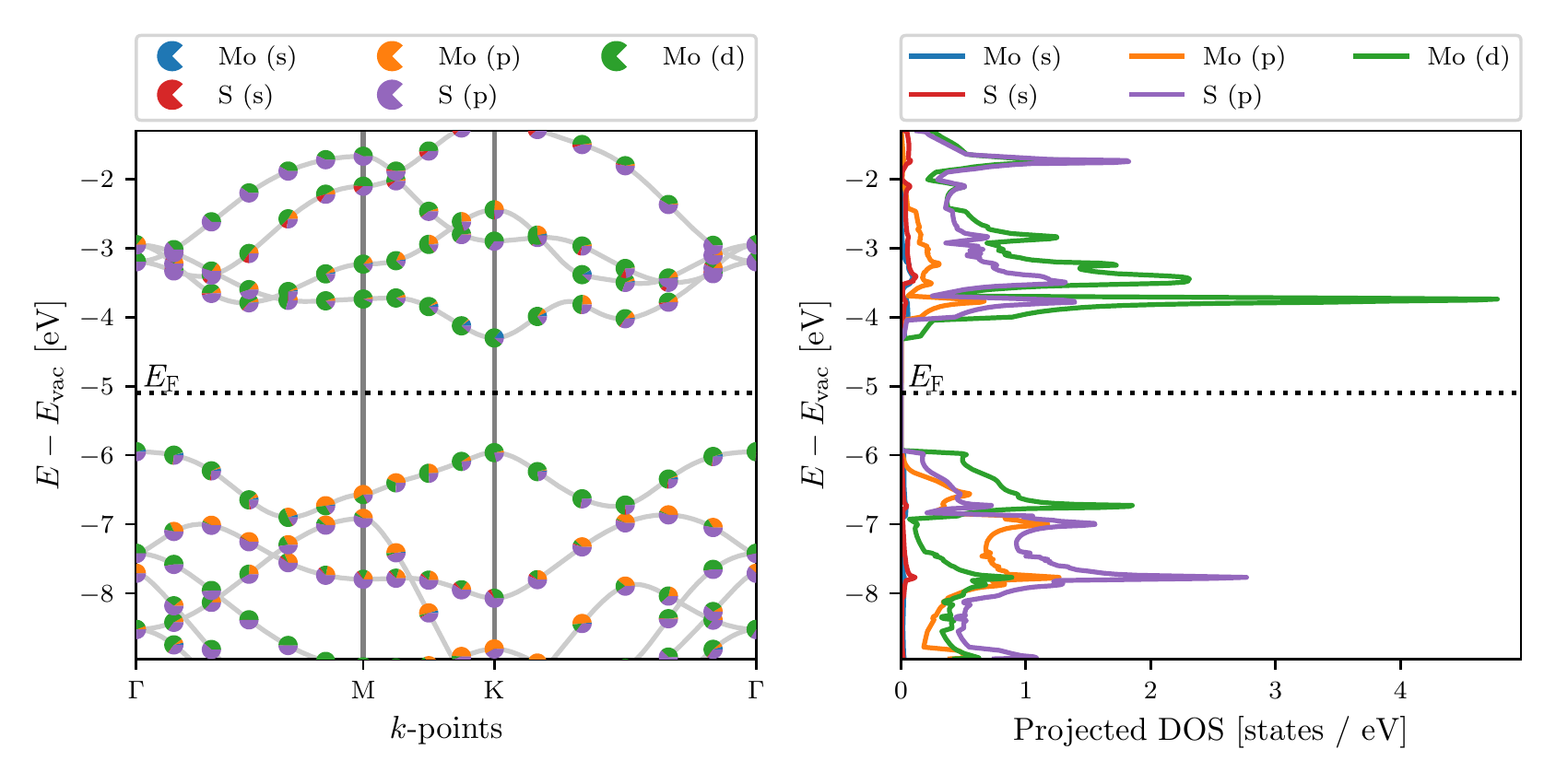}
    \caption{Orbital projected band structure and orbital projected density of states of $\textrm{MoS}_2$ in the H-phase. The pie chart symbols indicate the fractional atomic orbital character of the Kohn-Sham wave functions.}
    \label{fig:MoS2-H_pbands-pdos}
\end{figure*}
To facilitate a state-specific analysis of the PBE Kohn-Sham wave functions, an orbital projected band structure (PBS) is provided to complement the projected density of states (PDOS). In the PAW methodology, the all-electron wave functions are projected onto atomic orbitals inside the augmentation spheres centered at the position of each atom. The PBS resolves these atomic orbital contributions to the wave functions as a function of band and $k$-point whereas the PDOS resolves the atomic orbital character of the total density of states as a function of energy. The spin-orbit coupling is not included in the PBS or PDOS, as its effect is separately visualized by the spin-projected band structure also available in the C2DB. 

As an example, Figure \ref{fig:MoS2-H_pbands-pdos} shows the PBS (left) and PDOS (right) of monolayer $\textrm{MoS}_2$ calculated with PBE. The relative orbital contribution to a given Bloch state is indicated by a pie chart symbol. In the present example, one can deduce from the PBS that even though Mo-$p$ orbitals and S-$p$ orbitals contribute roughly equally to the DOS in the valence band, the Mo-$p$ orbital contributions are localized to a region in the BZ around the $M$-point, whereas the S-$p$ orbitals contribute throughout the entire BZ. 

%From the projected band structure it follows that the valence band states at $\Gamma$ have a significant contribution from S-$p$ orbitals as compared to the states at $K$, which are predominantly Mo-$d$. This explains the stronger interlayer hybridization observed at $\Gamma$ compared to K when MoS$_2$ layers are vertically stacked. The same conclusions apply to the conduction band minimum at K and the local minimum between $\Gamma-K$, where indeed the same hybridization behavior upon stacking is seen. While such state-specific analysis cannot be made on basis of the PDOS, we stress that the latter contains complementary information as it sums up the orbital character of states across the entire BZ, whereas the projected band structure is limited to a 1D path.

\subsection{Corrected G$_0$W$_0$ band structures}
\begin{figure}[t]
    \centering
    \includegraphics{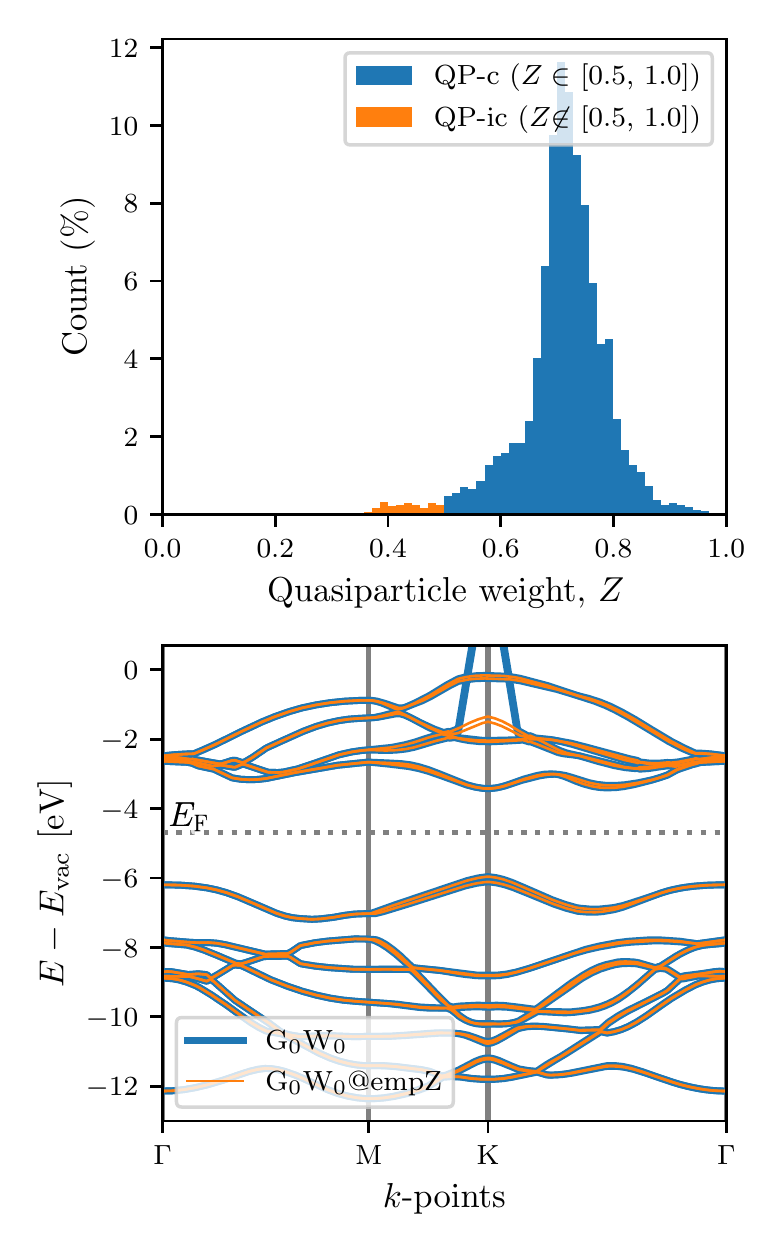}
    \caption{Top: Distribution of the 61716 QP weights ($Z$) contained in the C2DB. The blue part of the distribution shows QP-consistent (QP-c) $Z$-values while the orange part shows QP-inconsistent (QP-ic) $Z$ values. In general, the linear expansion of the self-energy performed when solving the QP equation works better for $Z$ closer to 1. About 0.3\% of the $Z$-values lie outside the interval from 0 to 1 and are not included in the distribution. Bottom: G$_0$W$_0$ band structure before (blue) and after (orange) applying the empZ correction, which replaces $Z$ by the mean of the distribution for QP-ic states. In the case of MoS$_2$ only one state at $K$ is QP-ic.}
    \label{fig:gwempzfigure}
\end{figure}

The C2DB contains \GW quasiparticle (QP) band structures of 370 monolayers covering 14 different crystal structures and 52 chemical elements. The details of these calculations can be found in the original C2DB paper\cite{haastrup2018computational}. A recent in-depth analysis of the 61.716 \GW data points making up the QP band structures led to several important conclusions relevant for high-throughput \GW calculations. In particular, it identified the linear QP approximation as a significant error source in standard \GW calculations and proposed an extremely simple correction scheme (the \textit{empirical Z} (empZ) scheme), that reduces this error by a factor of two on average. 

The empZ scheme divides the electronic states into two classes according to the size of the QP weight, $Z$. States with $Z \in [0.5, 1.0]$ are classified as QP consistent (QP-c) while states with $Z \not \in [0.5, 1.0]$ are classified as QP inconsistent (QP-ic). With this definition, QP-c states will have at least half of their spectral weight in the QP peak. The distribution of the 60.000+ $Z$-values is shown in Figure \ref{fig:gwempzfigure}. It turns out that the linear approximation to the self-energy, which is the gist of the QP approximation, introduces significantly larger errors for QP-ic states than for QP-c states. Consequently, the empZ method replaces the calculated $Z$ of QP-ic states with the mean of the $Z$-distribution, $Z_0 \approx 0.75$. This simple replacement reduces the average error of the linear approximation from 0.11 eV to 0.06 eV. 
 
An illustration of the method applied to MoS$_2$ is shown in Figure \ref{fig:gwempzfigure}. The original uncorrected \GW band structure is shown in blue while the empZ corrected band structure is shown in orange. MoS$_2$ has only one QP-ic state in the third conduction band at the $K$-point. Due to a break-down of the QP approximation for this state, the \GW correction is greatly overestimated leading to a local discontinuity in the band structure. The replacement of $Z$ by $Z_0$ for this particular state resolves the problem. All \GW band structures in the C2DB are now empZ corrected. 

\subsection{Optical absorbance}

In the first version of the C2DB, the optical absorbance was obtained from the simple expression \cite{haastrup2018computational} 
\begin{align}
\label{eq:absorbance1}
A(\omega) \approx \frac{\omega\Im{\alpha^{\mathrm{2D}}(\omega)}}{\epsilon_0c} \, ,
\end{align}
where $\alpha^{\mathrm{2D}}$ is the long wavelength limit of the in-plane sheet polarisability density (Note that the equation is written here in SI units). The sheet polarisability is related to the sheet conductivity via $\sigma^{\mathrm{2D}}(\omega)=-i\omega \alpha^{\mathrm{2D}}(\omega)$. The expression (\ref{eq:absorbance1}) assumes that the electric field inside the layer equals the incoming field (i.e. reflection is ignored), and hence, it may overestimate the absorbance. 

In the new version, the absorbance is evaluated from $A=1-R-T$, where $R$ and $T$ are the reflected and transmitted powers of a plane wave at normal incidence, respectively. These can be obtained from the conventional transfer matrix method applied to a monolayer suspended in vacuum. The 2D material is here modelled as an infinitely thin layer with a sheet conductivity. Alternatively, it can be modelled as quasi-2D material of thickness $d$ with a ``bulk'' conductivity of $\sigma = \sigma^{\mathrm{2D}}/d$ \cite{li2018two}, but the two approaches yield very similar results, since the optical thickness of a 2D material is much smaller than the optical wavelength. Within this model, the expression for the absorbance of a suspended monolayer with the sheet conductivity $\sigma^{\mathrm{2D}}$ reads
\begin{align}
\label{eq:absorbance}
A(\omega) = \Re{\sigma^{\mathrm{2D}}(\omega)\eta_0} \abs{\frac{2}{2+\sigma^{\mathrm{2D}}(\omega)\eta_0}}^2,
\end{align}
where $\eta_0=1/(\epsilon_0c) \approx 377$ $\Omega$ is the vacuum impedance. 

If the light-matter interaction is weak, i.e. $|\sigma^{\mathrm{2D}} \eta_0| \ll 1$, Eq. (\ref{eq:absorbance}) reduces to Eq.~(\ref{eq:absorbance1}). Nonetheless, due the strong light-matter interaction in some 2D materials, this approximation is not reliable in general. In fact, it can be shown that the maximum possible absorption from Eq.~(\ref{eq:absorbance}) is 50\%, which is known as the upper limit of light absorption in thin films \cite{hadley1947reflection}. This limit is not guaranteed by Eq.~(\ref{eq:absorbance1}), which can even yield an absorbance above 100\%. 

\begin{figure*}[t]
	\centering
	\includegraphics[width=0.9\textwidth]{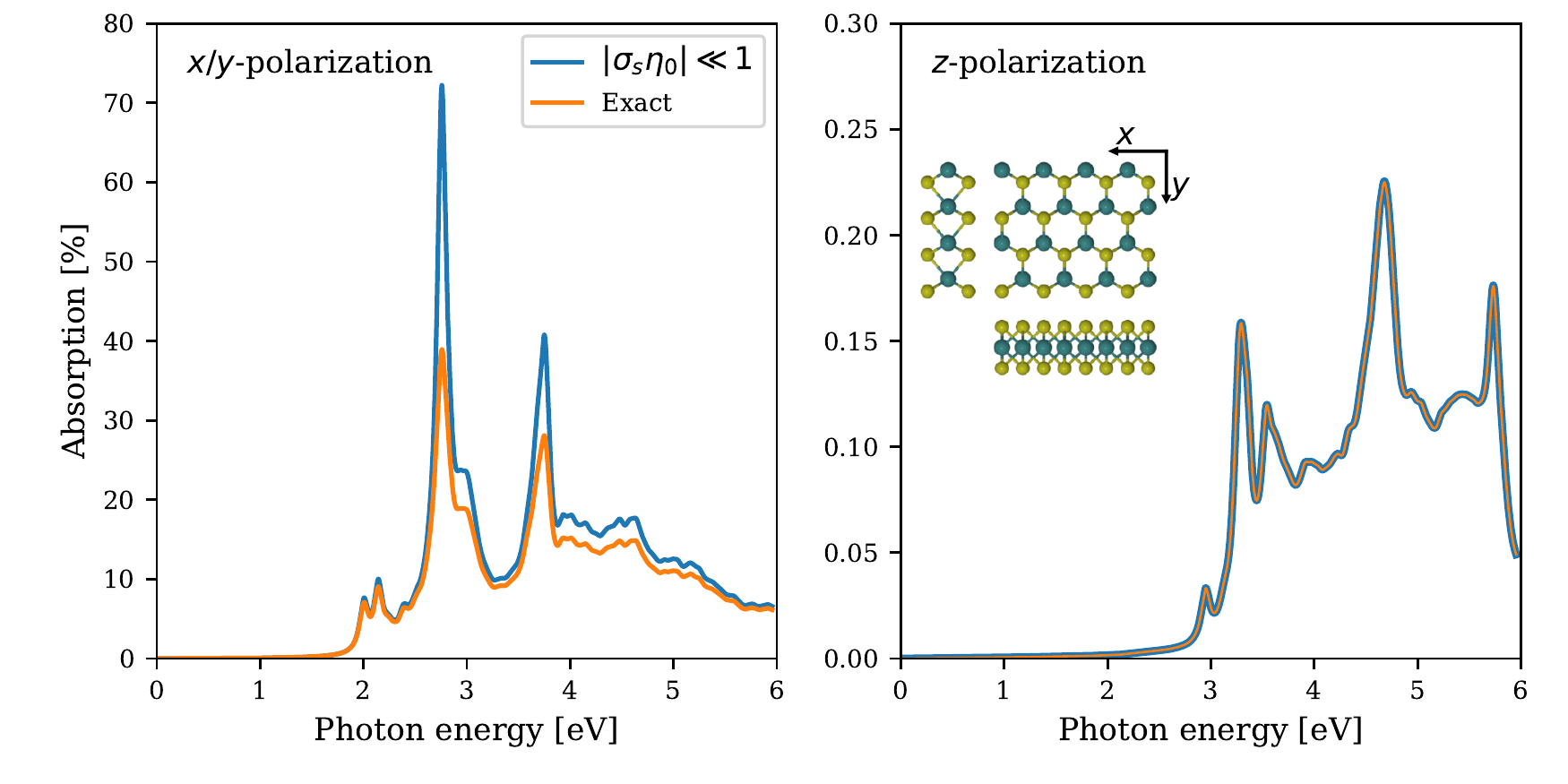}
	\caption[absorption]{Optical absorption of standalone monolayer MoS$_2$ for $x/y$-polarization (left) and $z$-polarization (right) at normal incident in the BSE framework, obtained using Eq.~(\ref{eq:absorbance1}) (blue) or Eq.~(\ref{eq:absorbance}) (orange). The crystal structure cross-sectional views are shown in the inset with the definition of directions.} 
	\label{fig:absorption}
\end{figure*}

As an example, Fig.~\ref{fig:absorption} shows the absorption spectrum of monolayer MoS$_2$ for in- and out-of-plane polarized light as calculated with the exact Eq.~(\ref{eq:absorbance}) and the approximate Eq.~(\ref{eq:absorbance1}), respectively. In all cases the sheet polarisability is obtained from the Bethe-Salpeter Equation (BSE) to account for excitonic effects \cite{haastrup2018computational}. For weak light-matter interactions, e.g. for the $z$-polarized light, the two approaches agree quite well, but noticeable differences are observed in regions with stronger light-matter interaction.

\section{New materials in the C2DB}\label{sec:newmat}
In this section we discuss the most significant extensions of the C2DB in terms of new materials. The set of materials presented here is not complete, but represents the most important and/or well defined classes. The materials discussed in Secs. \ref{sec:janus} and \ref{sec:experimental} (MXY Janus monolayers and monolayers extracted from experimental crystal structure databases) are already included in the C2DB. The materials described in Secs. \ref{sec:stacking} and \ref{sec:defect} (homo-bilayers and monolayer point defect systems) will soon become available as separate C2DB-interlinked databases. 

\subsection{MXY Janus monolayers}\label{sec:janus}
\begin{figure}[t]
    \centering
    \includegraphics[width=.9\linewidth]{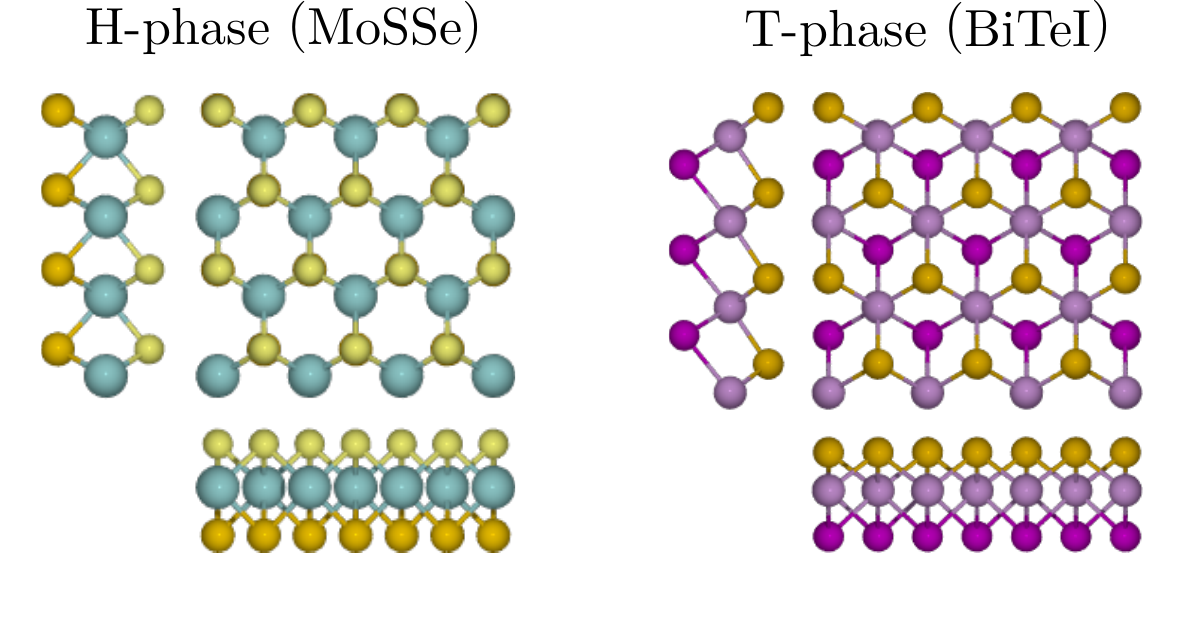}
    \caption{
      Atomic structure of the MXY Janus monolayers in the H-phase (left) and T-phase (right). The two prototype materials MoSSe and BiTeI are examples of experimentally realized monolayers adopting these crystal structures (not to scale).
    }
    \label{fig:janus_struc}
\end{figure}

The class of transition metal dichalcogenide (TMDC) monolayers
of the type MX$_2$ (where M is the transition metal and X is a chalcogen) exhibits a large variety of interesting and unique properties and has been widely discussed in the literature \cite{RevExTMDC}.
Recent experiments have shown that it is not only possible to synthesize different materials by changing the metal M or the chalcogen X, but also by exchanging the X on one side of the layer by another chalcogen (or halogen) \cite{lu_janus_2017,zhang2017janus,fulop2018exfoliation}.
This results in a class of 2D materials known as MXY Janus monolayers with broken mirror symmetry and finite out-of-plane dipole moments. The prototypical MXY crystal structures are shown in  Fig.~\ref{fig:janus_struc} for the case of MoSSe and BiTeI, which have both been experimentally realized \cite{lu_janus_2017,zhang2017janus,fulop2018exfoliation}. Adopting the nomenclature from the TMDCs, the crystal structures are denoted as H- or T-phase, depending on whether X and Y atoms are vertically aligned or displaced, respectively.

In a recent work \cite{c2dbJanus}, the C2DB workflow was employed to scrutinize and classify the basic electronic and optical properties of 224 different MXY Janus monolayers. All data from the study is available in the C2DB. Here we provide a brief discussion of the Rashba physics in these materials and refer the interested reader to Ref.~\cite{c2dbJanus} for more details and analysis of other properties.

A key issue when considering hypothetical materials, i.e. materials not previously synthesized, is their stability. The experimentally synthesized MoSSe and BiTeI are both found to be dynamically stable and lie within 10 meV of the convex hull confirming their thermodynamic stability. Out of the 224 initial monolayers 93 are classified as stable according to the C2DB criteria (dynamically stable and $\Delta H_\text{hull} < 0.2$\,eV/atom). Out of the 93 stable materials, 70 exhibits a finite band gap when computed with the PBE xc-functional.

\begin{figure*}[bt]
    \centering
    \includegraphics[width=\textwidth]{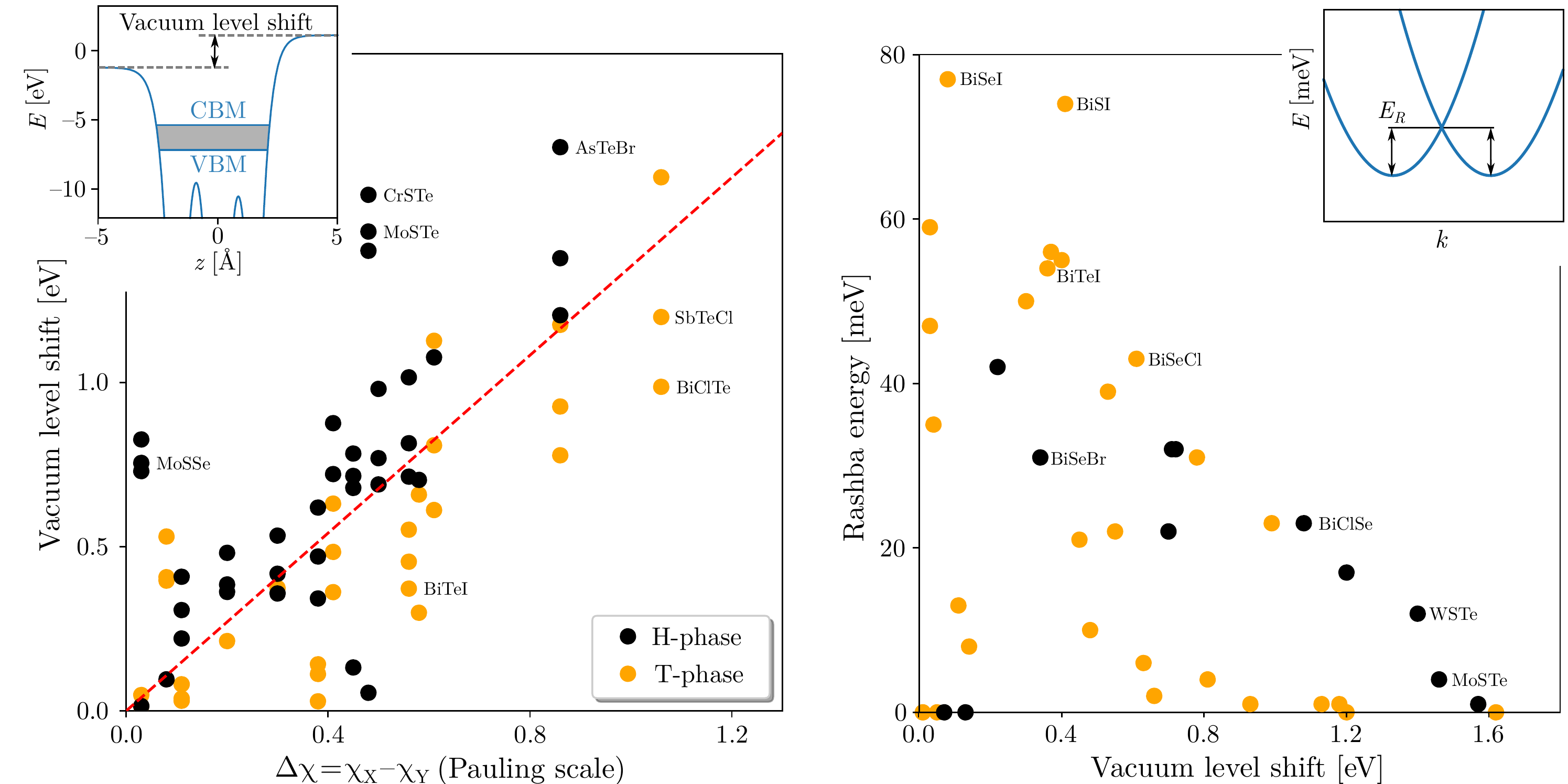}
    \caption{
      Left: Correlation between the electronegativity difference of X and Y in MXY Janus monolayers
      and the vacuum level shift across the layer. Right: Correlation between the Rashba energy and the vacuum level shift.
      Structures in the H-phase (e.g. MoSSe) are shown in black while structures in the T-phase (e.g. BiTeI) are shown in orange.
      The linear fit has a slope $1.35\,\mathrm{eV}/\Delta\chi$ (Pauling scale).
      The insets show the definition of the vacuum level shift and the Rashba energy, respectively.
      Modified from \cite{c2dbJanus}.
    }
    \label{fig:neg-dip}
\end{figure*}

The Rashba effect is a momentum dependent splitting of the band energies of a 2D semiconductor in the vicinity of a band extremum arising due to the combined effect of spin-orbit interactions and a broken crystal symmetry in the direction perpendicular to the 2D plane. The simplest model used to describe the Rashba effect is a 2D electron gas in a perpendicular electric field (along the $z$-axis). Close to the band extremum, the energy of the two spin bands is described by the Rashba Hamiltonian \cite{bychkov1984oscillatory,petersen2000simple}
\begin{align}
  H = \alpha_R (\boldsymbol{\sigma} \cross {\mathbf k})\cdot  \hat {\mathbf e}_z \, ,
\end{align}
where $\boldsymbol \sigma$ is the vector of Pauli matrices, $ {\mathbf k} = {\mathbf p} / \hbar$ is the wave number, and the Rashba parameter is proportional to the electric field strength, $\alpha_R \propto E_0$,  

Although the Rashba Hamiltonian is only meant as a qualitative model, it is of interest to test its validity on the Janus monolayers. The electric field of the Rashba model is approximately given by $E_0=\Delta V_{\mathrm{vac}}/d$, where $\Delta V_{\mathrm{vac}}$ is the shift in vacuum potential on the two sides of the layer (see left inset of Fig.~\ref{fig:neg-dip}) and $d$ is the layer thickness. Assuming a similar thickness for all monolayers, the electric field is proportional to the potential shift. Not unexpected, the latter is found to correlate strongly with the difference in electronegativity of the X and Y atoms, see left panel of Fig.~\ref{fig:neg-dip}.

% \alpha_R = 2 E_R / k_R
The Rashba energy, $E_R$, can be found 
by fitting $E(k) = \hbar^2 k^2 / 2m^* \pm \alpha_R k$ to the band structure (see right inset of Fig.~\ref{fig:neg-dip}) and should scale with the electric field strength. However, as seen from the right panel of Fig.~\ref{fig:neg-dip}, there is no correlation between the two quantities. Hence we conclude that the simple Rashba model is completely inadequate and that the strength of the perpendicular electric field cannot be used to quantify the effect of spin-orbit interactions on band energies. 

\subsection{Monolayers from known layered bulk crystals}\label{sec:experimental}
The C2DB has been extended with a number of monolayers that are likely exfoliable from experimentally known layered bulk compounds. Specifically, the Inorganic Crystal Structure Database (ICSD)\cite{Allmann2007} and Crystallography Open Database (COD)\cite{Grazulis2012} have first been filtered for corrupted, duplicate and theoretical compounds, which reduce the initial set of 585.485 database entries to 167.767 unique materials. All of these have subsequently been assigned a "dimensionality score" based on a purely geometrical descriptor. If the 2D score is larger than the sum of 0D, 1D and 3D scores we regard the material as being exfoliable and we extract the individual 2D components that comprise the material (see also Sec. \ref{sec:dimensionality}). We refer to the original work on the method for details \cite{Larsen2019} and note that similar approaches were applied in Refs. \cite{ashton2017topology,mounet2018two} to identify potentially exfoliable monolayers from the ICSD and COD.

The search has been limited to bulk compounds containing less than 6 different elements and no rare earth elements. This reduces the set of relevant bulk materials to 2991. For all of these we extracted the 2D components containing less than 21 atoms in the unit cell, which were then relaxed and sorted for duplicates following the general C2DB workflow steps described in Secs. \ref{sec:relax}-\ref{sec:similar}. At this point 781 materials remain. This set includes most known 2D materials and 207 of the 781 were already present in the C2DB prior to this addition. All the materials (including those that were already in C2DB) have been assigned an ICSD/COD identifier that refers to the parent bulk compound from which the 2D material was computationally exfoliated. We emphasize that we have not considered exfoliation energies in the analysis and a subset of these materials may thus be rather strongly bound and challenging to exfoliate even if the geometries indicate van der Waals bonded structures of the parent bulk compounds.
\begin{figure}[t]
    \centering
    \includegraphics[width=0.45\textwidth]{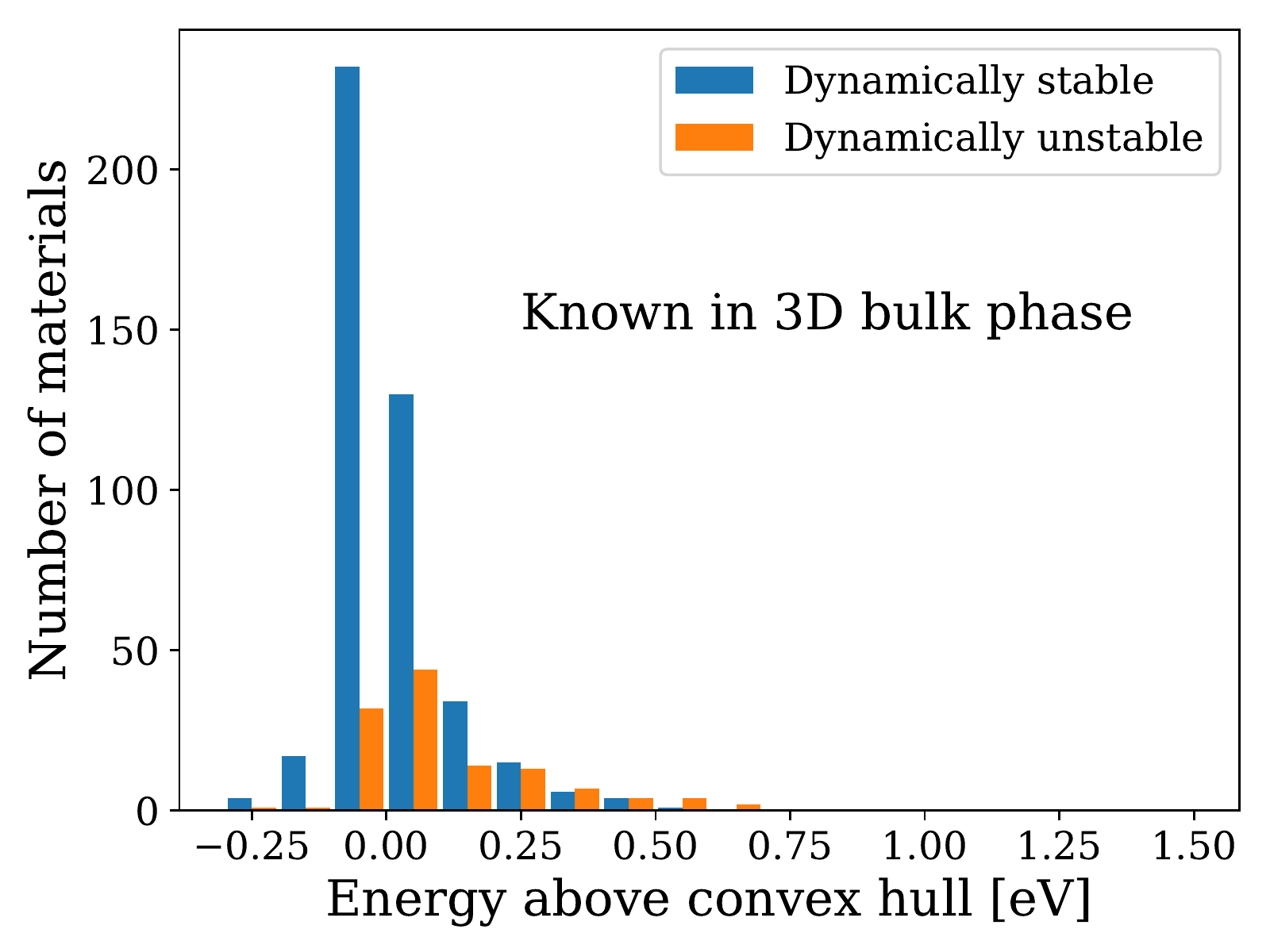}
    \includegraphics[width=0.45\textwidth]{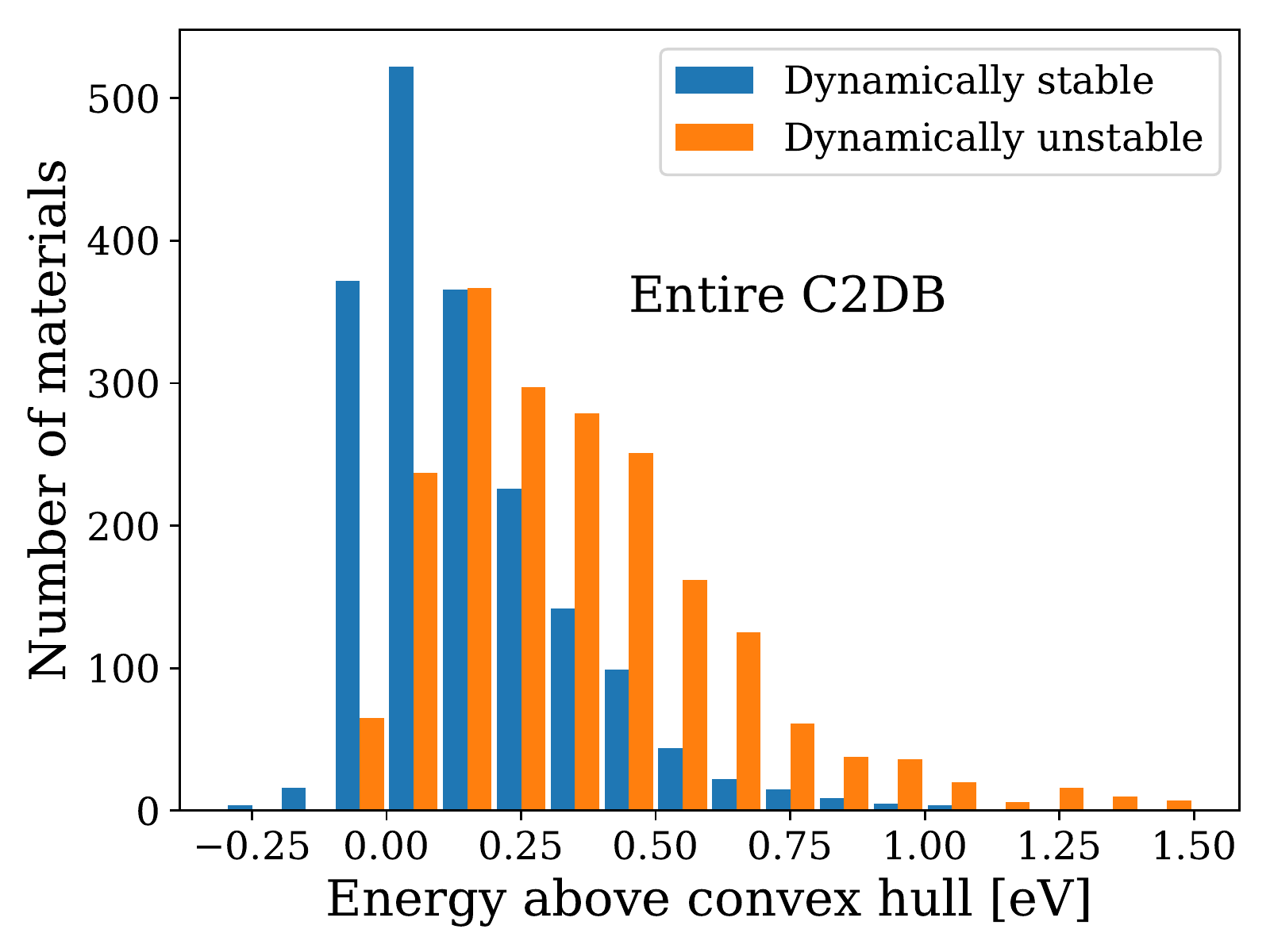}
    \caption{Distribution of energies above the convex hull for the 2D materials extracted from bulk compounds in ICSD and COD (top) and for the entire C2DB including those constructed from combinatorial lattice decoration (bottom). Dynamically stable materials are indicated in blue.}
    \label{fig:icsd_vs_c2db}
\end{figure}

Fig. \ref{fig:icsd_vs_c2db} shows the distribution of energies above the convex hull for materials derived from parent structures in ICSD or COD as well as for the entire C2DB, which includes materials obtained from combinatorial lattice decoration as well. As expected, the materials derived from experimental bulk materials are situated rather close to the convex hull whereas those obtained from lattice decoration extend to energies far above the convex hull. It is also observed that a larger fraction of the experimentally derived materials are dynamically stable. There are, however, well known examples of van der Waals bonded structures where the monolayer undergoes a significant lattice distortion, which will manifest itself as a dynamical instability in the present context. For example, bulk MoS$_2$ exists in van der Waals bonded structures composed of either 2H-MoS$_2$ or 1T-MoS$_2$ layers, but a monolayer of the 1T phase undergoes a structural deformation involving a doubling of the unit cell\cite{Qian2014} and is thus categorized as dynamically unstable by the C2DB workflow. The dynamically stable materials derived from parent bulk structures in the ICSD and COD may serve as a useful subset of the C2DB that are likely to be exfoliable from known compounds and thus facilitate experimental verification. As a first application the subset has been used to search for magnetic 2D materials, which resulted in a total of 85 ferromagnets and 61 anti-ferromagnets \cite{Torelli2020}.

\subsection{Outlook: Multilayers}\label{sec:stacking}
\begin{figure}
    \centering
    \includegraphics{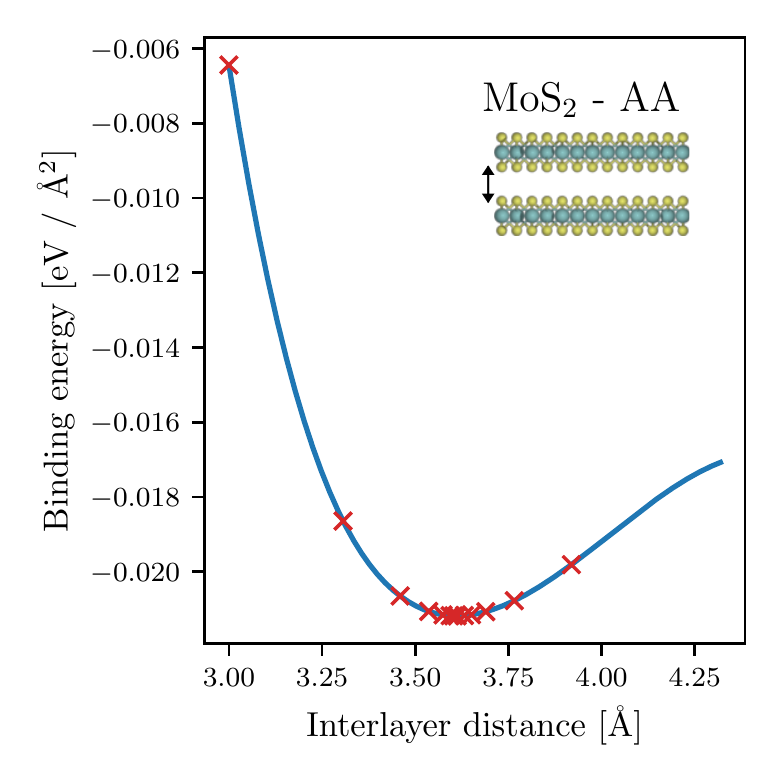}
    \caption{An illustration of the optimization of the interlayer (IL) distance for MoS$_2$ in the AA stacking. The black crosses are the points sampled by the optimization algorithm while the blue curve is a spline interpolation of the black crosses. The inset shows the MoS$_2$ AA stacking and the definition of the IL distance is indicated with a black double-sided arrow.}
    \label{fig:mos2binding}
\end{figure}
\begin{figure}
    \centering
    \includegraphics{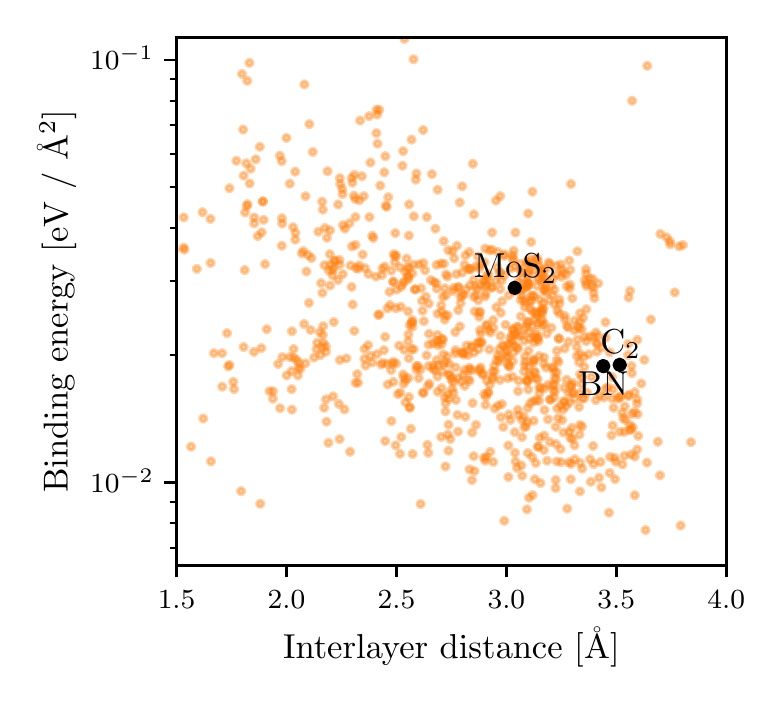}
    \caption{Scatter plot of the calculated interlayer distance and binding energies of (homo)bilayers of selected materials from C2DB. A few well known materials are highlighted: MoS$_2$, graphene (C$_2$), and hexagonal boron-nitride (BN). The bilayer binding energies provide an estimate of the monolayer exfoliation energies, see Sec. \ref{sec:exfoliation}}
    \label{fig:scatterbindings}
\end{figure}

The C2DB is concerned with the properties of covalently bonded monolayers (see discussion of dimensionality filtering in Sec.~\ref{sec:dimensionality}). However, multilayer structures composed of two or more identical monolayers are equally interesting and often have properties that deviate from those of the monolayer. In fact, the synthesis of layered vdW structures with a controllable number of layers represents an interesting avenue for atomic-scale materials design. Several examples of novel phenomena emerging in layered vdW structures have been demonstrated including direct-indirect band gap transitions in MoS2$_2$\cite{mak2010atomically,splendiani2010emerging}, layer-parity selective Berry curvatures in few-layer WTe$_2$\cite{xiao2020berry}, thickness-dependent magnetic order in CrI$_3$\cite{sivadas2018stacking,liu2019thickness}, and emergent ferroelectricity in bilayer hBN\cite{yasuda2020stacking}.

As a first step towards a systematic exploration of multilayer 2D structures, the C2DB has been used as basis for generating homobilayers in various stacking configurations and subsequently computing their properties following a modified version of the C2DB monolayer workflow. Specifically, the most stable monolayers (around 1000) are combined into bilayers by applying all possible transformations (unit cell preserving point group operations and translations) of one layer while keeping the other fixed. The candidate bilayers generated in this way are subject to a stability analysis, which evaluates the binding energy and optimal interlayer distance based on PBE-D3 total energy calculations keeping the atoms of the monolayers fixed in their PBE relaxed geometry, see Fig. \ref{fig:mos2binding} and Table \ref{tab:exfoliationenergy}.

The calculated interlayer binding energies are generally in the range from a few to a hundred meV/\AA$^2$ and interlayer distances range from 1.5\AA\ to 3.8\AA. A scatter plot of preliminary binding energies and interlayer distances is shown in Fig. \ref{fig:scatterbindings}. The analysis of homobilayers provides an estimate of the energy required to peel a monolayer off a bulk structure. In particular, the binding energy for the most stable bilayer configuration provides a measure of the \emph{exfoliation energy} of the monolayer. This key quantity is now available for all monolayers in the C2DB, see Sec. {\ref{sec:exfoliation}}.

\subsection{Outlook: Point defects}\label{sec:defect}
\begin{figure*}[t]
    \centering
    \includegraphics[width=1\textwidth]{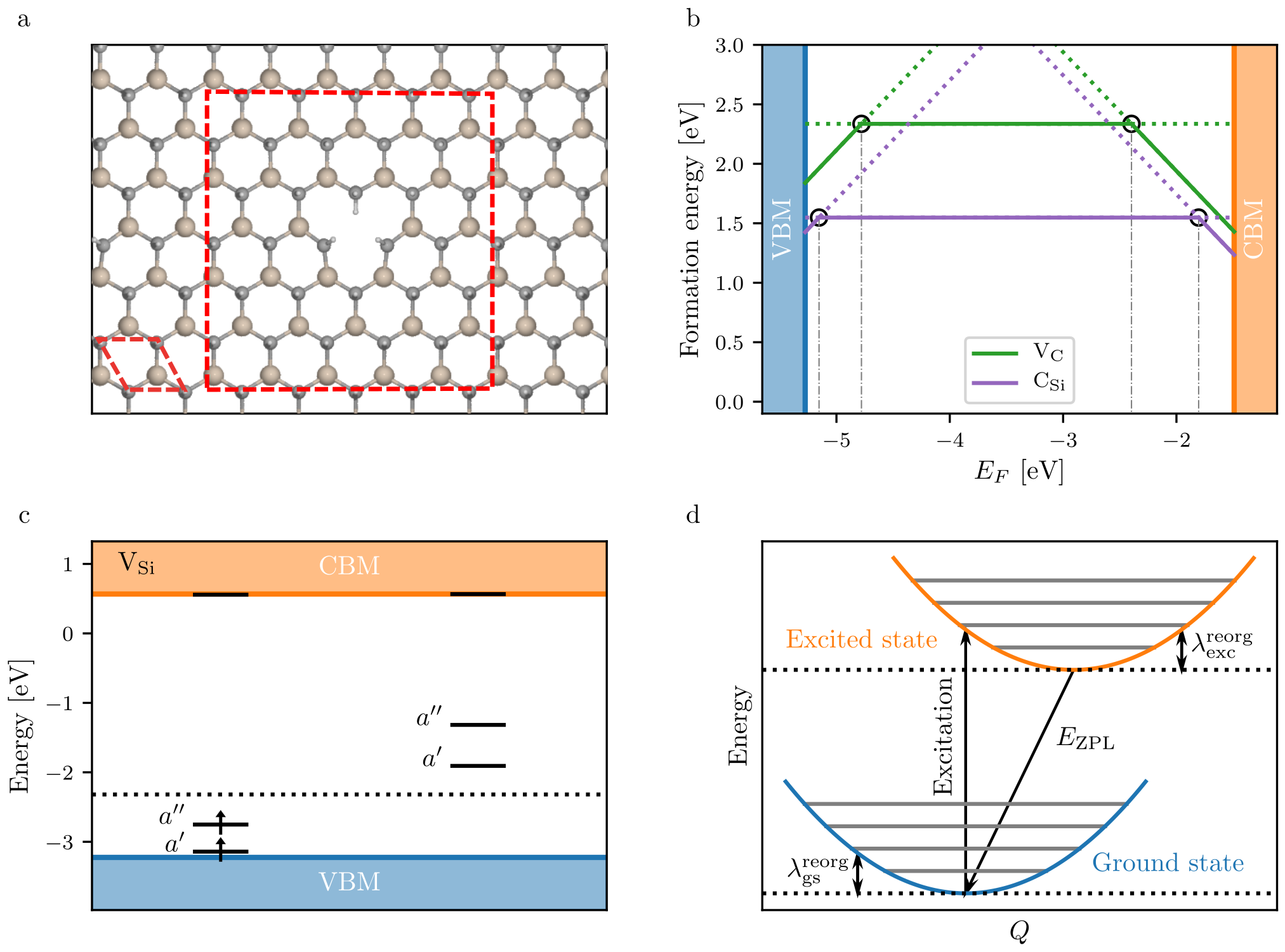}
    \caption{Overview of some of the properties included in the 2D defect database project for the example host material CH$_2$Si. (a) The supercell used to represent the defects (here a Si vacancy). The supercell is deliberately chosen to break the symmetry of the host crystal lattice. (b) Formation energies of a C vacancy (green) and C-Si substitutional defect (purple). (c) Energy and orbital symmetry of the localized single-particle states of the  $\mathrm{V_{Si}}$ defect for both spin channels (left and right). The Fermi level is shown by the dotted line. (d) Schematic excited state configuration energy diagram. The transitions corresponding to the vertical absorption and the zero-phonon emission are indicated.}
    \label{fig:defects}
\end{figure*}
The C2DB is concerned with the properties of 2D materials in their pristine crystalline form. However, as is well known the perfect crystal is an idealized model of real materials, which always contain defects in smaller or larger amounts depending on the intrinsic materials properties and growth conditions. Crystal defects often have a negative impact on physical properties, e.g. they lead to scattering and life time-reduction of charge carriers in semiconductors. However, there are also important situations where defects play a positive enabling role, e.g. in doping of semiconductors, as color centers for photon emission\cite{northup2014quantum,o2009photonic} or as active sites in catalysis.\par
%The influence of defects in semiconductors has already been studied for various systems as their behaviour can be highly important for technological applications \textbf{[ADD CITATIONS!]}. For instance, single-photon emitters (SPEs) play an increasingly central role in quantum technologies like quantum communication and quantum information processing \cite{northup2014quantum,o2009photonic}. From a experimental point of view, it is much easier to embed and control defects in 2D systems compared to bulk materials. However, a systematic study of various point defects over a large set of 2D semiconductors is yet to be conducted.\par
To reduce the gap between the pristine model material and real experimentally accessible samples, a systematic evaluation of the basic properties of the simplest native point defects in a selected subset of monolayers from the C2DB has been initiated. The monolayers are selected based on the stability of the pristine crystal. Moreover, only non-magnetic semiconductors with a PBE band gap satisfying $E_{\mathrm{gap}} > 1$ eV, are currently considered as such materials are candidates for quantum technology applications like single-photon sources and spin qubits. Following these selection criteria around 300 monolayers are identified and their vacancies and intrinsic substitutional defects are considered, yielding a total of about 1500 defect systems.\par  

Each defect system is subject to the same workflow, which is briefly outlined below. To enable point defects to relax into their lowest energy configuration, the symmetry of the pristine host crystal is intentionally broken by the chosen supercell, see Fig. \ref{fig:defects} (a). In order to minimize defect-defect interaction, supercells are furthermore chosen such that the minimum distance between periodic images of defects is larger than $15$ \AA. Unique point defects are created based on the analysis of equivalent Wyckoff positions for the host material. To illustrate some of the properties that will feature in the upcoming point defect database, we consider the specific example of monolayer CH$_2$Si.\par
First, the formation energy\cite{zhang1991chemical,van1993first} of a given defect is calculated from PBE total energies. Next, Slater-Janak transition state theory is used to obtain the charge transition levels\cite{janak1978proof,pandey2016defect}. By combining these results, one obtains the formation energy of the defect in all possible charge states as a function of the Fermi level. An example of such a diagram is shown in Fig. \ref{fig:defects} (b) for the case of the $\mathrm{V_{C}}$ and $\mathrm{C_{Si}}$ defects in monolayer CH$_2$Si. For each defect and each charge state, the PBE single-particle energy level diagram is calculated to provide a qualitative overview of the electronic structure. A symmetry analysis\cite{kaappa2018point} is performed for the defect structure and the individual defect states lying inside the band gap. The energy level diagram of the neutral $\mathrm{V_{Si}}$ defect in CH$_2$Si is shown in Fig. \ref{fig:defects} (c), where the defect states are labeled according to the irreducible representations of the $C_\mathrm{s}$ point group.\par 
In general, excited electronic states can be modelled by solving the Kohn-Sham equations with non-Aufbau occupations. The excited-state solutions are saddle points of the Kohn-Sham energy functional, but common self-consistent field (SCF) approaches often struggle to find such solutions, especially when nearly degenerate states are involved. The calculation of excited states corresponding to transitions between localized states inside the band gap is therefore performed using an alternative method based on the direct optimization (DO) of orbital rotations in combination with the maximum overlap method (MOM) \cite{levi2020variational}. This method ensures fast and robust convergence of the excited states, as compared to SCF. In Fig. \ref{fig:defects} (d), the reorganization energies for the ground- and excited state, as well as the zero-phonon line (ZPL) energy are sketched. For the specific case of the Si vacancy in CH$_2$Si, the DO-MOM method yields $E_{\mathrm{ZPL}} = 3.84$ eV, $\lambda_{\mathrm{gs}}^{\mathrm{reorg}} = 0.11$ eV and $\lambda_{\mathrm{exc}}^{\mathrm{reorg}} = 0.16$ eV.  For systems with large electron-phonon coupling (i.e. Huang-Rhys factor $>1$) a one-dimensional approximation for displacements along the main phonon mode is used to produce the configuration coordinate diagram (see Fig. \ref{fig:defects} (d)). In addition to the ZPL energies and reorganization energies, the Huang-Rhys factors, photoluminescence spectrum from the 1D phonon model, hyperfine coupling and zero field splitting are calculated.

\section{New properties in the C2DB}\label{sec:newprop}
This section reports on new properties that have become available in the C2DB since the first release. The employed computational methodology is described in some detail and results are compared to the literature where relevant. In addition, some interesting property correlations are considered along with general discussions of the general significance and potential application of the available data.

\begin{table}[t!]
\centering
%\begin{tabular}{ m{2cm}  m{3cm}  m{3cm} }
\begin{tabular}{lllll}
\hline
Material & SG & PBE+D3 & DF2 & rVV10 \\
\hline\hline
MoS$_2$ & P-6m2 & 28.9 & 21.6 & 28.8 \\
MoTe$_2$ & P-6m2 & 30.3 & 25.2 & 30.4 \\
ZrNBr & Pmmn& 18.5 & 10.5 & 18.5 \\
C & P6/mmm & 18.9 & 20.3 & 25.5 \\
P & Pmna & 21.9 & 38.4 & 30.7 \\
BN & P-6m2 & 18.9 & 19.4 & 24.4 \\
WTe$_2$ & P-6m2 & 32.0 & 24.7 & 30.0 \\
PbTe & P3m1 & 23.2 & 27.5 & 33.0 \\
\hline
\end{tabular}
\caption{Exfoliation energies for selected materials calculated with the PBE+D3 xc-functional as described in Sec. \ref{sec:stacking} and compared with the DF2 and rVV10 results from Ref. \cite{mounet2018two}. The spacegroups are indicated in the column "SG". All numbers are in units of meV/Å$^2$. \label{tab:exfoliationenergy}}
\end{table}

\subsection{Exfoliation energy}\label{sec:exfoliation}
The exfoliation energy of a monolayer is estimated as the binding energy of its bilayer in the most stable stacking configuration (see also Sec. \ref{sec:stacking}). The binding energy is calculated using the PBE+D3 xc-functional\cite{grimme2010consistent} with the atoms of both monolayers fixed in the PBE relaxed geometry. Table \ref{tab:exfoliationenergy} compares exfoliation energies obtained in this way to values from Mounet \emph{et al.}\cite{mounet2018two} for a representative set of monolayers.    

\subsection{Bader charges}\label{sec:bader}
For all monolayers we calculate the net charge on the individual atoms using the Bader partitioning scheme\cite{Bader}. The analysis is based purely on the electron density, which we calculate from the PAW pseudo density plus compensation charges using the PBE xc-functional. Details of the method and its implementation can be found in Tang \textit{et al.}\cite{Tang_2009}. In Sec. \ref{sec:born} we compare and discuss the relation between Bader charges and Born charges.  

\subsection{Spontaneous polarization}
The spontaneous polarization ($\mathbf{P}_\mathrm{s}$) of a bulk material is defined as the charge displacement with respect to that of a reference centrosymmetric structure \cite{Resta1992,KingSmithVanderbilt1993}. Ferroelectric materials exhibit a finite value of $\mathbf{P}_\mathrm{s}$ that may be switched by an applied external field
and have attracted a large interest for a wide range of applications \cite{zhang2011piezoelectric,maeder2004lead,Scott2000}.

\begin{figure}[t]
	\centering
	\includegraphics{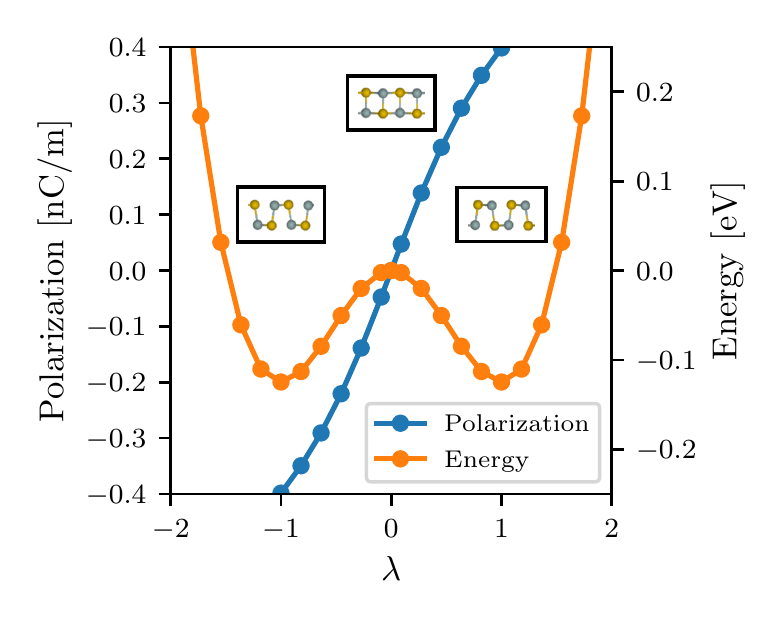}
	\caption{Depicted in the blue plot is the formal polarization calculated along the adiabatic path for GeSe, using the methods described in the main text. The orange plot shows the energy potential along the path as well as outside. Figure inset: The structure of GeSe in the 2 non-centrosymmetric configurations corresponding to -$\mathbf{P}_\mathrm{s}$ and $\mathbf{P}_\mathrm{s}$ and the centrosymmetric configuration.} 
	\label{fig:polarization_figure}
\end{figure}

The spontaneous polarization in bulk materials can be regarded as electric dipole moment per unit volume, but in contrast to the case of finite systems this quantity is ill-defined for periodic crystals \cite{Resta1992}. Nevertheless, one can define the formal polarization density
\begin{align}
\mathbf{P} = \frac{1}{2\pi}\frac{e}{V}\sum_l\phi_l\mathbf{a}_l
\label{eq:formal_polarization}
\end{align}
where $\mathbf{a}_l$ (with $l\in \lbrace 1,2,3 \rbrace $) are the lattice vectors spanning the unit cell, $V$ is the cell volume and $e$ is the elementary charge.  $\phi_l$ is the polarization phase along the lattice vector defined by
\begin{align}
\phi_l = \sum_{i}Z_{i}\mathbf{b}_l\cdot\mathbf{u}_{i} - \phi_l^\mathrm{elec}
\label{eq:polarization_phase}
\end{align}
where $\mathbf{b}_l$ is the reciprocal lattice vector satisfying $\mathbf{b}_l\cdot\mathbf{R}_l=2\pi$ and  $\mathbf{u}_{i}$ is the position of nucleus $i$ with charge $eZ_i$. The electronic contribution to the polarization phase is defined as
\begin{align}
    \phi_l^\mathrm{elec} =&
    \frac{1}{N_{k\perp \mathbf{b}_l}}\Im\sum_{k\in \mathrm{BZ}_{\perp\mathbf{b}_l}}
    \nonumber\\ &\times \ln \prod_{j=0}^{N_{k\parallel\mathbf{b}_l}-1}
    \det_{occ} \left[\bra{u_{n\mathbf{k} + j \delta\mathbf{k}}}\ket{ u_{m\mathbf{k}+(j + 1)\delta\mathbf{k}}}\right],
\end{align}
where $\mathrm{BZ}_{\perp \mathbf{b}_l} = \lbrace {\mathbf{k} | \mathbf{k}\cdot\mathbf{b}_l = 0} \rbrace$ is a plane of $\mathbf{k}$-points orthogonal to $\mathbf{b}_l$, $\delta\mathbf{k}$ is the distance between neighbouring k-points in the $\mathbf{b}_l$ direction and $N_{k\parallel\mathbf{b}_l}$ ($N_{k\perp\mathbf{b}_l}$) is the number of $\mathbf{k}$-points along (perpendicular to) the $\mathbf{b}_l$ direction. These expression generalize straightforwardly to 2D.

The formal polarization is only well-defined modulo $e\mathbf{R}_n/V$ where $\mathbf{R}_n$ is any lattice vector. However, changes in polarization are well defined and the spontaneous polarization may thus be obtained by 
\begin{align}\label{eq:adiabatic_path}
    \mathbf{P}_\mathrm{s}=\int_0^1\frac{d\mathbf{P}(\lambda)}{d\lambda}d\lambda,
\end{align}
where $\lambda$ is a dimensionless parameter that defines an adiabatic structural path connecting the polar phase ($\lambda=1$) with a non-polar phase ($\lambda=0$).

The methodology has been implemented in GPAW and used to calculate the spontaneous polarization of all stable materials in the C2DB with a PBE band gap above 0.01 eV and a polar space group symmetry. For each material, the centrosymmetric phase with smallest atomic displacement from the polar phase is constructed and relaxed under the constraint of inversion symmetry. The adiabatic path connecting the two phases is then used to calculate the spontaneous polarization using Eqs.~\eqref{eq:formal_polarization}-\eqref{eq:adiabatic_path}. An example of a calculation for GeSe is shown in Fig. \ref{fig:polarization_figure} where the polarization along the path connecting two equivalent polar phases via the centrosymmetric phase is shown together with the total energy. The spontaneous polarization obtained from the path is $39.8$ nC/m in good agreement with previous calculations \cite{polarization_2D_monochalcogenides}.

\subsection{Born charges}\label{sec:born}

The Born charge of an atom $a$ at position $\mathbf u_a$ in a solid is defined as
\begin{align}
    Z^a_{ij} = \frac{V}{e} \frac{\partial P_i}{\partial u_{aj}}\Bigg|_{E=0}\label{eq:born} \, .
\end{align}
It can be understood as an effective charge assigned to the atom to match the change in polarization in direction $i$ when its position is perturbed in direction $j$. Since the polarization density and the atomic position are both vectors, the Born charge of an atom is a rank-2 tensor. The Born charge is calculated as a finite difference and relies on the Modern theory of polarization \cite{Resta2007} for the calculation of polarization densities, see Ref. \cite{Gjerding2020} for more details. The Born charge has been calculated for all stable materials in C2DB with a finite PBE band gap.

It is of interest to examine the relation between the Born charge and the Bader charge (see Sec. \ref{sec:bader}). In materials with strong ionic bonds one would expect the charges to follow the atoms. On the other hand, in covalently bonded materials the hybridization pattern and thus the charge distribution, depends on the atom positions in a complex way, and the idea of charges following the atom is expected to break down. In agreement with this idea, the (in-plane) Born charges in the strongly ionic hexagonal boron-nitride ($\pm 2.71 e$ for B and N, respectively) are in good agreement with the calculated Bader charges ($\pm 3.0 e$). In contrast, (the in-plane) Born charges in MoS$_2$ ($-1.08 e$ and $0.54 e$ for Mo and S, respectively) deviate significantly from the Bader charges ($1.22 e$ and $-0.61 e$ for Mo and S, respectively). In fact, the values disagree even on the sign of the charges underlining the non-intuitive nature of the Born charges in covalently bonded materials. 

Note that the out-of-plane Born charges never match the Bader charges, even for strongly ionic insulators, and are consistently smaller in value than the in-plane components. The smaller out-of-plane values are consistent with the generally smaller out-of-plane polarisability of 2D materials (for both electronic and phonon contributions) and agrees with the intuitive expectation that it is more difficult to polarize a 2D material in the out-of-plane direction as compared to the in-plane direction. 

\begin{figure}
    \centering
    \includegraphics{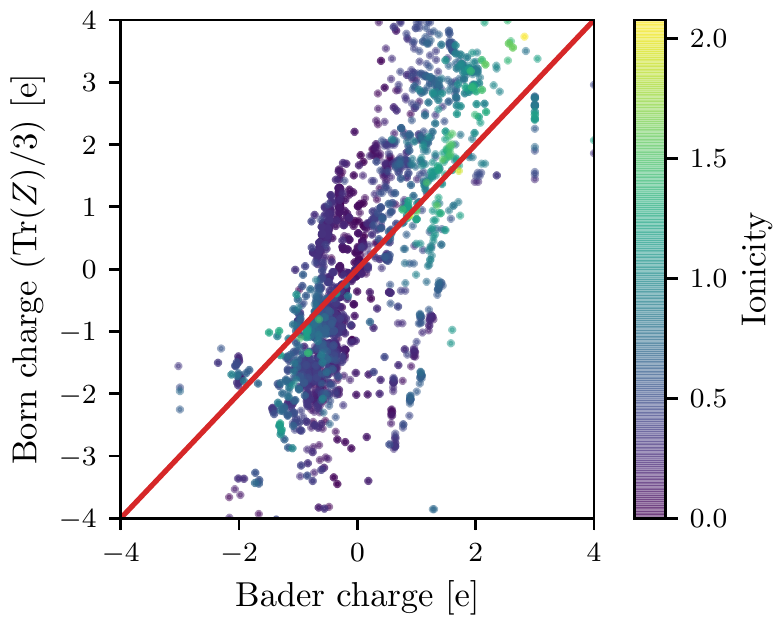}
    \caption{Born charges, $\mathrm{Tr}(Z)/3$, vs. Bader charges for 3025 atoms in the 585 materials for which the Born charges are calculated.  The colors indicate the ionicity of the atoms (see main text).}
    \label{fig:bader-born}
\end{figure}

Fig. \ref{fig:bader-born} shows the average of the diagonal of the Born charge tensor, $\mathrm{Tr}(Z^a)/3$, plotted against the Bader charges for all 585 materials in the C2DB for which the Born charges have been computed.
The data points have been colored according to the ionicity of the atom $a$ defined as $I(a) = | \chi_a - \langle\chi\rangle |$, 
where $\chi_a$ and $\langle \chi \rangle $ are the Pauling electronegativity of atom $a$ and the average electronegativity of all atoms in the unit cell, respectively.
The ionicity is thus a measure of the tendency of an atom to donate/accept charge relative to the average tendency of atoms in the material. It is clear from Fig. \ref{fig:bader-born} that there is a larger propensity for the Born and Bader charges to match in materials with higher ionicity. 

\begin{figure}
    \centering
    \includegraphics{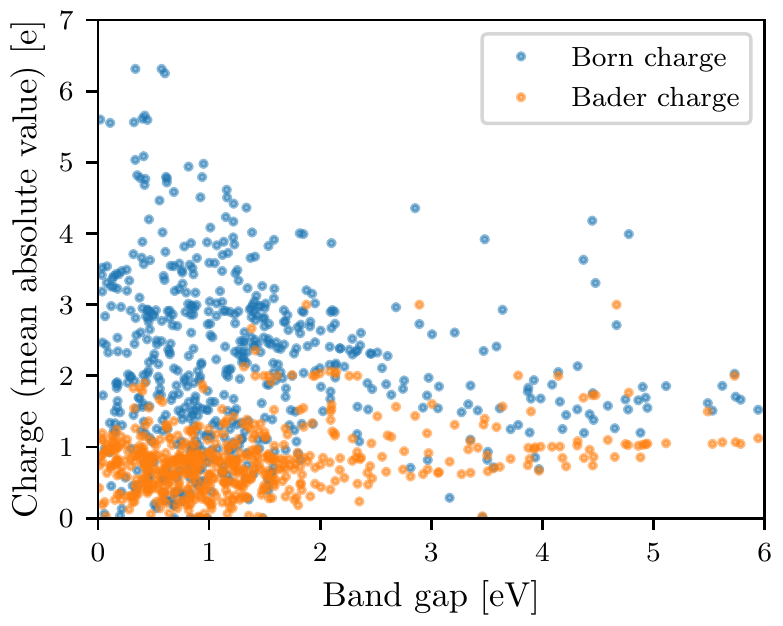}
    \caption{Bader and in-plane Born charges vs. band gap.}
    \label{fig:charge-gap}
\end{figure}

Fig. \ref{fig:charge-gap} plots the average (in-plane) Born charge and the Bader charge versus the band gap. It is clear that large band gap materials typically exhibit integer Bader charges, whereas there is no clear correlation between the Born charge and the band gap.


\subsection{Infrared polarizability}
\begin{figure}
    \centering
    \includegraphics{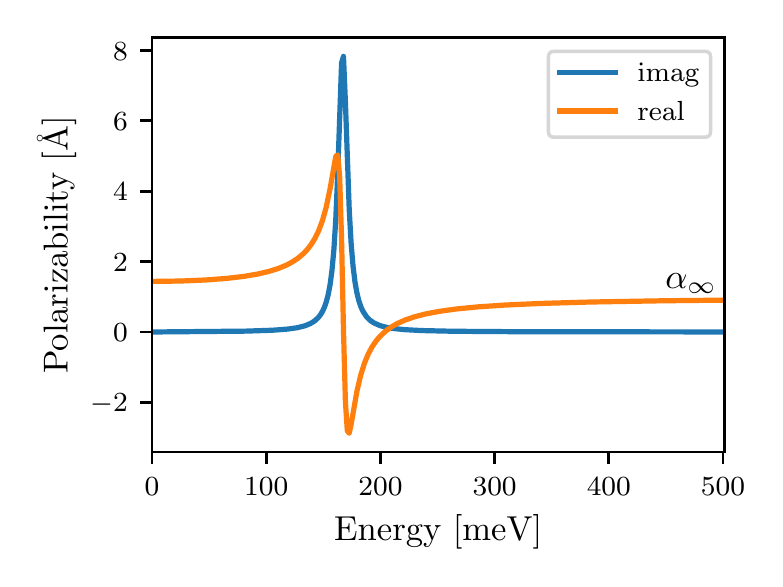}
    \caption{Total polarizability, including both electrons and phonons, of monolayer hBN in the infrared frequency regime. The resonance at around 180 meV is due to the $\Gamma$-point longitudinal optical phonon. At energies above all phonon frequencies (but below the band gap) the polarizability is approximately constant and equal to the static limit of the electronic polarizability, $\alpha_{\infty}$.}
    \label{fig:irpol}
\end{figure}

The original C2DB provided the frequency dependent polarisability computed in the random phase approximation (RPA) with inclusion of electronic interband and intraband (for metals) transitions\cite{haastrup2018computational}. However, phonons carrying a dipole moment (so-called infrared (IR) active phonons) also contribute to the  polarizability at frequencies comparable to the frequency of optical phonons. This response is described by the IR polarizability, 

\begin{align}
        \alpha^\mathrm{IR}(\omega) &= \frac{e^2}{A} \mathbf{Z}^T \mathbf{M}^{-1/2} \left(\sum_i\frac{\mathbf{d}_i\mathbf{d}_i^T}{\omega^2_i-\omega^2-i\gamma\omega}\right) \mathbf{M}^{-1/2} \mathbf{Z},
\end{align}
where $\mathbf{Z}$ and $\mathbf{M}$ are matrix representations of the Born charges and atomic masses, $\omega^2_i$ and $d_i$ are eigenvectors and eigenvalues of the dynamical matrix, $A$ is the in-plane cell area and $\gamma$ is a broadening parameter representing the phonon lifetime and is set to $10$ meV. The total polarizability is then the sum of the electronic polarizability and the IR polariability. 

The new C2DB includes the IR polarisability of all monolayers for which the Born charges have been calculated (stable materials with a finite band gap), see Sec. (\ref{sec:born}). As an example, Fig. \ref{fig:irpol} shows the total polarizability of monolayer hexagonal boron nitride. For details on the calculation of the IR polarizability see Ref. \cite{Gjerding2020}.

\subsection{Piezoelectric tensor}

The piezoelectric effect is the accumulation of charges, or equivalently the formation of an electric polarisation, in a material in response to an applied mechanical stress or strain. It is an important material characteristic with numerous scientific and technological applications in sonar, microphones, accelerometers, ultrasonic transducers, energy conversion etc \cite{ye2008handbook, ogawa2016piezoelectric}. The change in polarization originates from the movement of positive and negative charge centers as the material is deformed.  

Piezoelectricity can be described by the (proper) piezoelectric tensor $c_{ijk}$ with $i,j,k \in \lbrace x,y,z \rbrace$, given by \cite{Vanderbilt1999} 
\begin{align}
\label{eq:piezoelectric}
c_{ijk} = \frac{e}{2\pi V} \sum_{l} \frac{\partial\phi_l}{\partial \epsilon_{jk}} a_{li} \, .
\end{align}
which differs from Eq. (\ref{eq:formal_polarization}) only by a derivative of the polarization phase with respect to the strain tensor $\epsilon_{jk}$. Note that $c_{ijk}$ does not depend on the chosen branch cut.

The piezoelectric tensor is a symmetric tensor with at most 18 independent components. Furthermore, the point group symmetry restricts the number of independent tensor elements and their relationships due to the well-known Neumann's principle \cite{authier2003international}. For example, monolayer MoS$_2$ with point group $D_{3h}$, has only one non-vanishing independent element of $c_{ijk}$. Note that $c_{ijk}$ vanishes identically for centrosymmetric materials. Using a finite-difference technique with a finite but small strain (1\% in our case), Eq.~(\ref{eq:piezoelectric}) has been used to compute the proper piezoelectric tensor for all non-centrosymmetric materials in the C2DB with a finite band gap.  
Table~\ref{tab:piezoelectric} shows a comparison of the piezoelectric tensors in the C2DB with literature for a selected set of monolayer materials. Good agreement is obtained for all these materials.

\begin{table}[t!]
\centering
%\begin{tabular}{ m{2cm}  m{3cm}  m{3cm} }
\begin{tabular}{lllll}
\hline
Material &Exp. & Theory \cite{Duerloo2012} & C2DB \\ 
\hline\hline
BN & - & 0.14 & 0.13 \\
MoS$_2$ & 0.3  & 0.36 & 0.35 \\
MoSe$_2$ & - & 0.39 & 0.38 \\
MoTe$_2$ & - & 0.54 & 0.48 \\
WS$_2$ & - & 0.25 & 0.24 \\
WSe$_2$ & - & 0.27 & 0.26 \\
WTe$_2$ & - & 0.34 & 0.34 \\
\hline
\end{tabular}
\caption{Comparison of computed piezoelectric tensor versus experimental values and previous calculations for hexagonal BN and a selected set of TMDs (space group 187). All number are in units of nC/m. Experimental data for MoS$_2$ is obtained from Ref.~\cite{Zhu2015}. \label{tab:piezoelectric}}
\end{table}

\subsection{Topological invariants}
For all materials in the C2DB exhibiting a direct band gap below 1 eV, the $k$-space Berry phase spectrum of the occupied bands has been calculated from the PBE wave functions. Specifically, a particular $k$-point is written as $k_1\mathbf{b}_1+k_2\mathbf{b}_2$ and the Berry phases $\gamma_n(k_2)$ of the occupied states on the path $k_1=0\rightarrow k_1=1$ is calculated for each value of $k_2$. The connectivity of the Berry phase spectrum determines the topological properties of the 2D Bloch Hamiltonian \cite{Taherinejad2014, Olsen2016a}.

The calculated Berry phase spectra of the relevant materials are available for visual inspection on the C2DB webpage. Three different topological invariants have been extracted from these spectra and are reported in the C2DB: 1) The Chern number, $C$, takes an integer value and is well defined for any gapped 2D material. It determines the number of chiral edge states on any edge of the material. For any non-magnetic material the Chern number vanishes due to time-reversal symmetry. It is determined from the Berry phase spectrum as the number of crossings at any horizontal line in the spectrum. 2) The mirror Chern number, $C_M$, defined for gapped materials with a mirror plane in the atomic layer\cite{fu2011topological}. For such materials, all states may be chosen as mirror eigenstates with eigenvalues $\pm i$ and the Chern numbers $C_{\pm}$ can be defined for each mirror sector separately. For a material with vanishing Chern number, the mirror Chern number is defined as $C_M=(C_+-C_-)/2$ and takes an integer value corresponding to the number of edge states on any mirror symmetry preserving edge. It is obtained from the Berry phase spectrum as the number of chiral crossings in each of the mirror sectors. 3) The $Z_2$ invariant, $\nu$, which can take the values 0 and 1, is defined for materials with time-reversal symmetry. Materials with $\nu=1$ are referred to as quantum spin Hall insulators and exhibit helical edge states at any time-reversal conserving edge. It is determined from the Berry phase spectrum as the number of crossing points modulus 2 at any horizontal line in the interval $k_2\in[0,1/2]$. 

\begin{figure*}[t]
    \centering
    \includegraphics[width=0.45\textwidth]{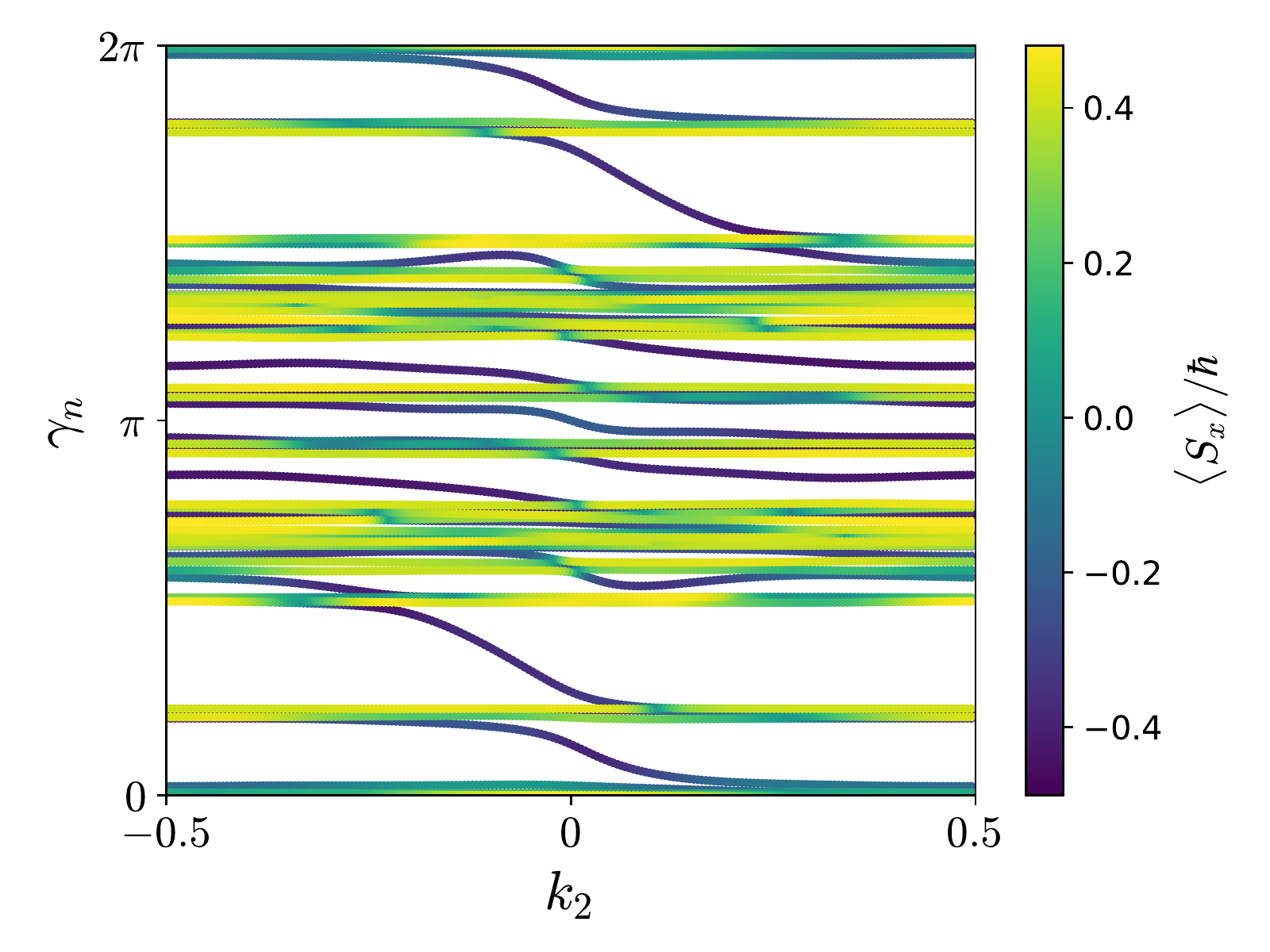}
    \includegraphics[width=0.45\textwidth]{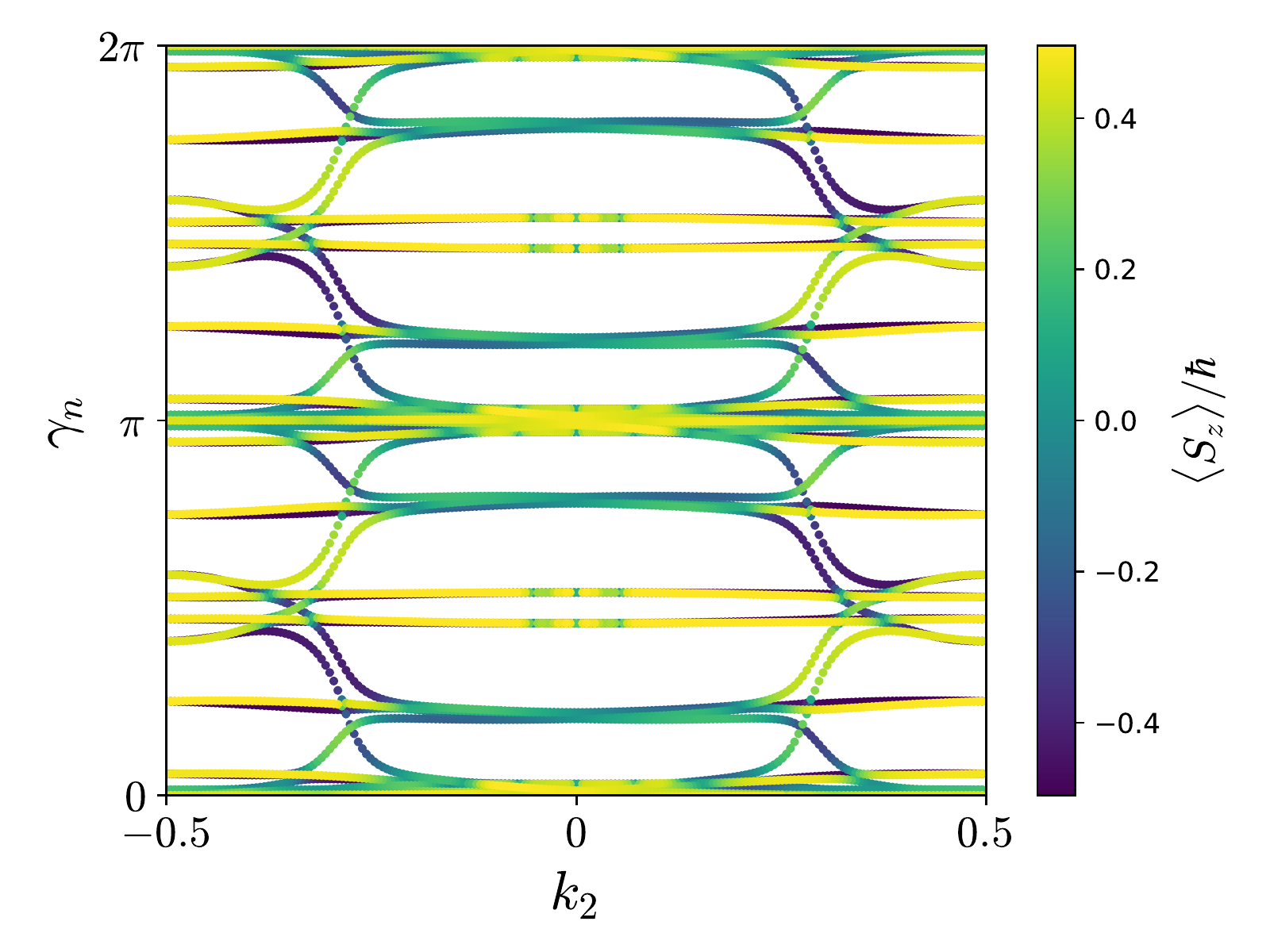}
    \includegraphics[width=0.45\textwidth]{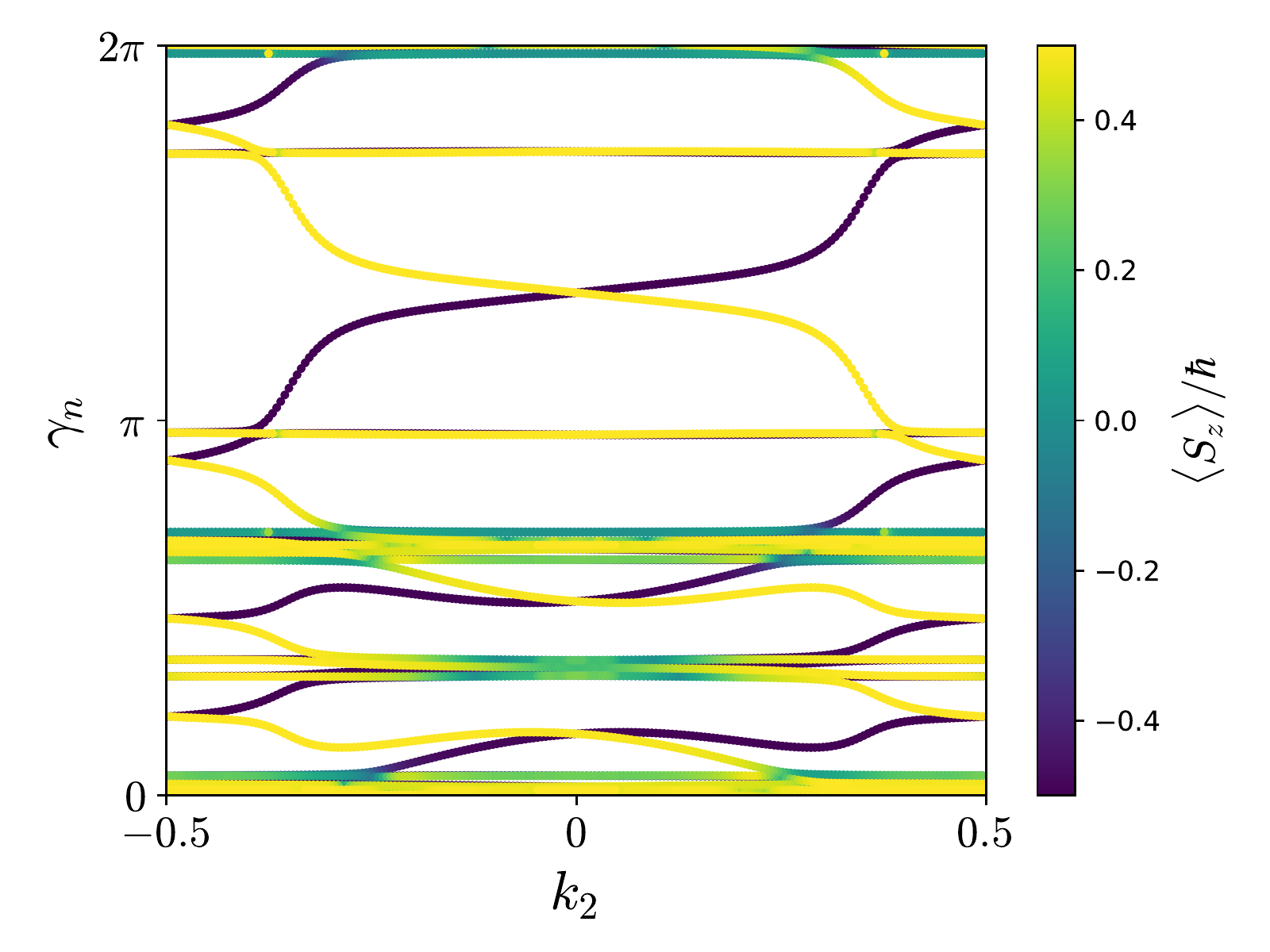}
    \includegraphics[width=0.45\textwidth]{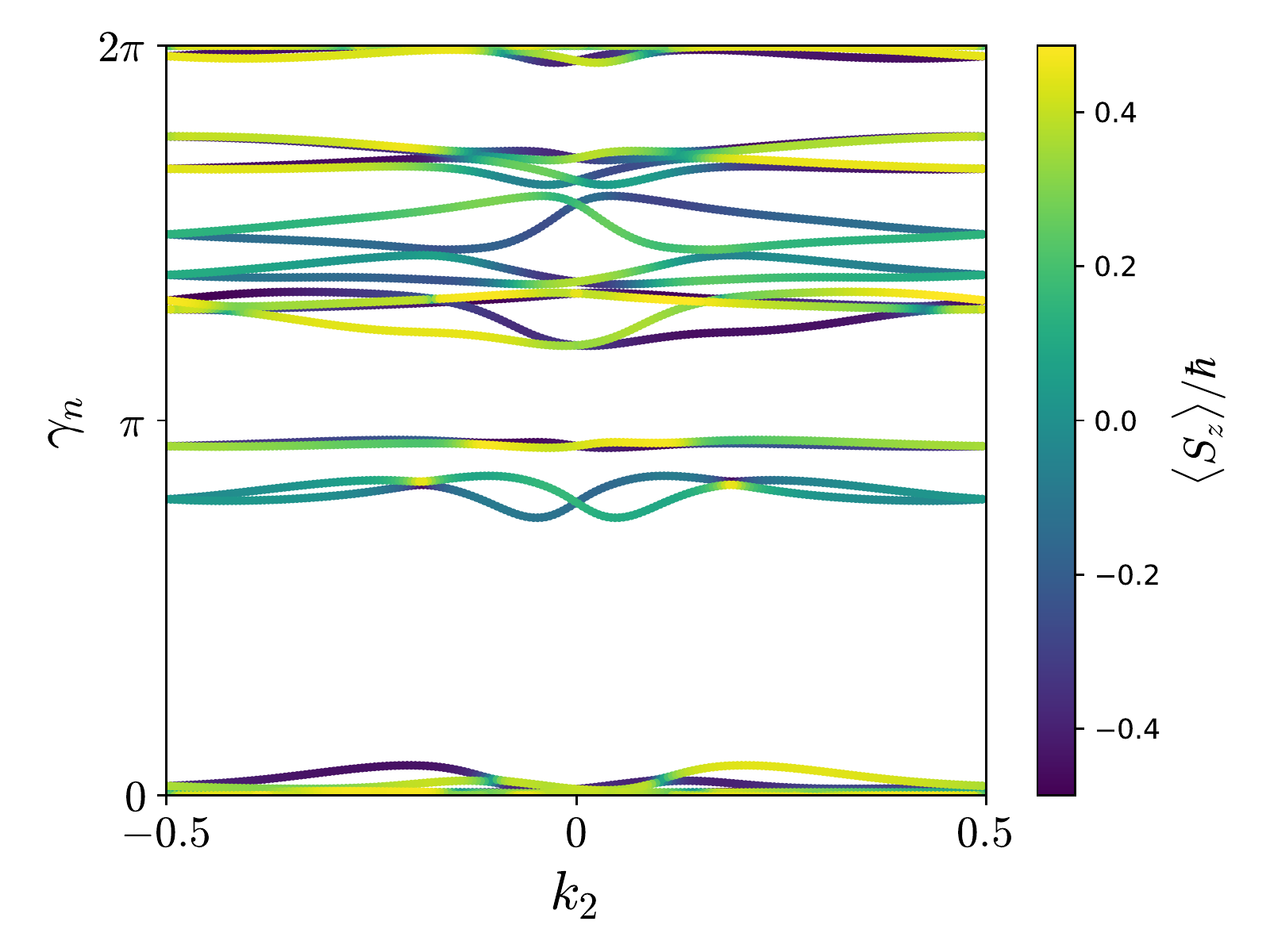}
    \caption{Berry phase spectra of the Chern insulator OsCl$_3$ (top left), the crystalline topological insulator OsTe$_2$ (top right), the quantum spin Hall insulator SbI (lower left) and the trivial insulator BiITe (lower right).}
    \label{fig:berry_phase}
\end{figure*}

Figure \ref{fig:berry_phase} shows four representative Berry phase spectra corresponding to the three cases of non-vanishing $C$, $C_M$ and $\nu$ as well as a trivial insulator. The four materials are: OsCl$_3$ (space group 147) - a Chern insulator with $C=1$, OsTe$_2$ (space group 14) - a mirror crystalline insulator with $C_M=2$, SbI (spacegroup 1) - a quantum spin Hall insulator with $\nu=1$ and BiITe (spacegroup 156) - a trivial insulator. Note that a gap in the Berry phase spectrum always implies a trivial insulator.

In Ref. \cite{Olsen2018} the C2DB was screened for materials with non-trivial topology. At that point it was found that the database contained 7 Chern insulators, 21 mirror crystalline topological insulators and 48 quantum spin Hall insulators. However, that does not completely exhaust the the topological properties of materials in the C2DB. In particular, there may be materials that can be topologically classified based on crystalline symmetries other than the mirror plane of the layer. In addition, second order topological effects may be present in certain materials, which imply that flakes will exhibit topologically protected corner states. Again, the Berry phase spectra may be used to unravel the second order topology by means of nested Wilson loops \cite{Benalcazar2017a}.

\subsection{Exchange coupling constants}\label{sec:exchange}

The general C2DB workflow described in Secs. \ref{sec:relax}-\ref{sec:similar} will identify the ferromagnetic ground state of a material and apply it as starting point for subsequent property calculations, whenever it is more stable than the spin-paired ground state. In reality, however, the ferromagnetic state is not guaranteed to comprise the magnetic ground state. In fact, anti-ferromagnetic states often have lower energy than the ferromagnetic one, but in general it is non-trivial to obtain the true magnetic ground state. We have chosen to focus on the ferromagnetic state due to its simplicity and because its atomic structure and stability are often very similar to those of other magnetic states. Whether or not the ferromagnetic state is the true magnetic ground state is indicated by the nearest neighbor exchange coupling constant as described below.

When investigating magnetic materials the thermodynamical properties (for example the critical temperatures for ordering) are of crucial interest. In two dimensions the Mermin-Wagner theorem\cite{MerminWagner} comprises an extreme example of the importance of thermal effects since it implies that magnetic order is only possible at $T=0$ unless the spin-rotational symmetry is explicitly broken. The thermodynamic properties cannot be accessed directly by DFT. Consequently, magnetic models that capture the crucial features of magnetic interactions must be employed. For insulators, the Heisenberg model has proven highly successful in describing magnetic properties of solids in 3D as well as 2D\cite{Olsen2019}. It represents the magnetic degrees of freedom as a lattice of localized spins that interact through a set of exchange coupling constants. 
If the model is restricted to include only nearest neighbor exchange and assume magnetic isotropy in the plane, it reads
\begin{align}\label{eq:heisenberg}
    H=-\frac{J}{2}\sum_{\langle ij\rangle}\mathbf{S}_i\cdot\mathbf{S}_j-\frac{\lambda}{2}\sum_{\langle ij\rangle}S_i^zS_j^z-A\sum_i\big(S_i^z\big)^2
\end{align}
where $J$ is the nearest neighbor exchange constant, $\lambda$ is the nearest neighbor anisotropic exchange constant and $A$ measures the strength of single-ion anisotropy. We also neglect off-diagonal exchange coupling constants that give rise to terms proportional to $S_i^xS_j^y$, $S_i^yS_j^z$ and $S_i^zS_j^x$. The out-of-plane direction has been chosen as $z$ and $\langle ij\rangle$ implies that for each site $i$ we sum over all nearest neighbor sites $j$. The parameters $J$, $\lambda$ and $A$ may be obtained from an energy mapping analysis involving four DFT calculations with different spin configurations\cite{Olsen2017, Torelli2020, Torelli2020a}. The thermodynamic properties of the resulting "first principles Heisenberg model" may subsequently be analysed with classical Monte Carlo simulations or renormalized spin wave theory \cite{Lado2017, Torelli2019}. 

The C2DB provides the values of $J$, $\lambda$, and $A$ as well as the number of nearest neighbors $N_{nn}$ and the maximum eigenvalue of $S_z$ ($S$), which is obtained from the total magnetic moment per atom in the ferromagnetic ground state (rounded to nearest half-integer for metals). These key parameters facilitate easy post-processing analysis of thermal effects on the magnetic structure. In Ref. \cite{Torelli2019b} such an analysis was applied to estimate the critical temperature of all ferromagnetic materials in the C2DB based on a model expression for $T_C$ and the parameters from Eq. \eqref{eq:heisenberg}. 

For metals, the Heisenberg parameters available in C2DB should be used with care because the Heisenberg model is not expected to provide an accurate description of magnetic interactions in this case. Nevertheless, even for metals the sign and magnitude of the parameters provide an important qualitative measure of the magnetic interactions that may be used to screen and select materials for more detailed investigations of magnetic properties. 

\begin{figure}[t]
    \centering
    \includegraphics[width=0.45\textwidth]{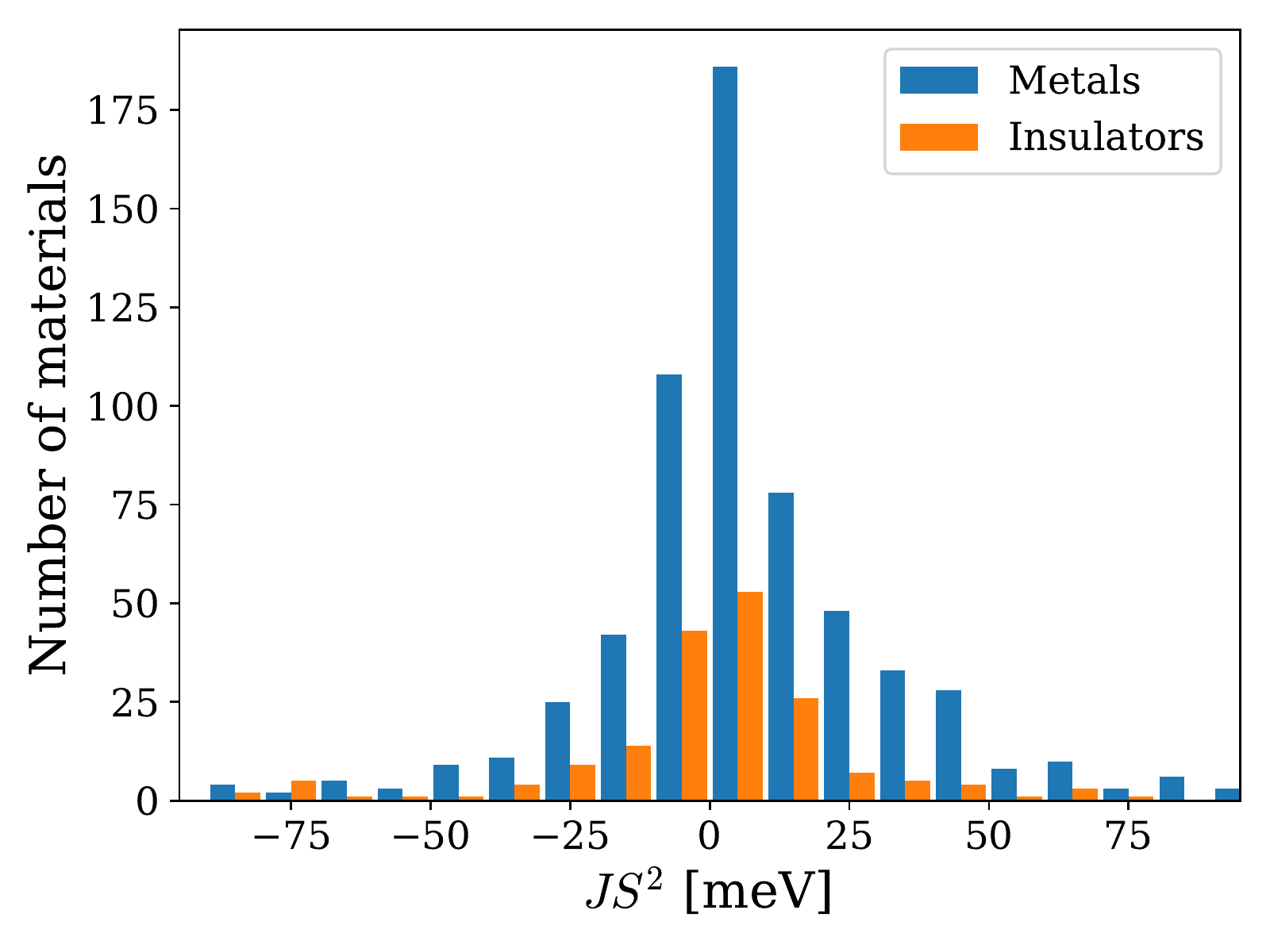}
    \includegraphics[width=0.45\textwidth]{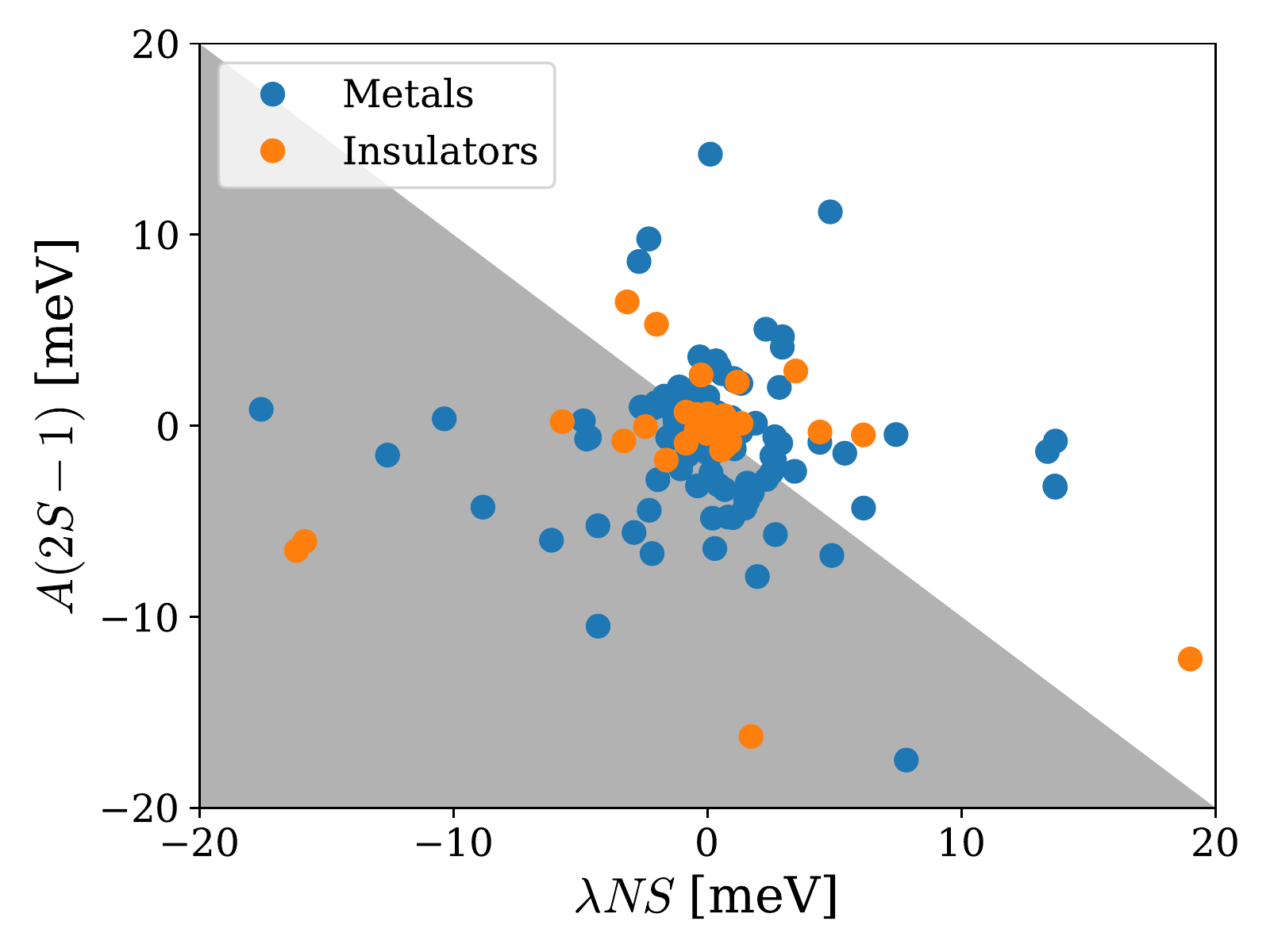}
    \caption{Top: Distribution of exchange coupling constants in C2DB. Bottom: Single-ion anisotropy $A$ vs anisotropic exchange $\lambda$ for ferromagnetic materials with $S>1/2$. The shaded area indicates the part of parameter space where the model (Eq.~\eqref{eq:heisenberg}) does not yield an ordered state at finite temperatures.}
    \label{fig:exchange_ani}
\end{figure}

A negative value of $J$ implies the existence of an anti-ferromagnetic state with lower energy than the ferromagnetic state used in C2DB. This parameter is thus crucial to consider when judging the stability and relevance of a material classified as magnetic in C2DB (see Sec. \ref{sec:magnetic}). Fig. \ref{fig:exchange_ani} shows the distribution of exchange coupling constants (weighted by $S^2$) of the magnetic materials in the C2DB. The distribution is slightly skewed to the positive side indicating that ferromagnetic order is more common than anti-ferromagnetic order.

The origin of magnetic anisotropy may stem from either single-ion anisotropy or anisotropic exchange and it is in general difficult {\it a priori} to determine, which mechanism is most important. There is, however, a tendency in the literature to neglect anisotropic exchange terms in a Heisenberg model description of magnetism and focus solely on the single-ion anisotropy. In Fig. \ref{fig:exchange_ani} we show a scatter plot of the anisotropy parameters $A$ and $\lambda$ for the ferromagnetic materials ($J>0$). The spread of the parameters indicate that the magnetic anisotropy is in general equally likely to originate from both mechanisms and neglecting anisotropic exchange is not advisable. For ferromagnets, the model (Eq.~\eqref{eq:heisenberg}) only exhibits magnetic order at finite temperatures if $A(2S-1)+\lambda N_{nn}>0$ \cite{Torelli2019b}. Neglecting anisotropic exchange thus excludes materials with $A<0$ that satisfies $A(2S-1)+\lambda N_{nn}>0$. This is in fact the case for 11 ferromagnetic insulators and 31 ferromagnetic metals in the C2DB.

\subsection{Raman spectrum}
Raman spectroscopy is an important technique used to probe the vibrational modes of a solid (or molecule) by means of inelastic scattering of light \cite{long2002raman}. In fact, Raman spectroscopy is the dominant method for characterising 2D materials and can yield detailed information about chemical composition, crystal structure and layer thickness. There exist several different types of Raman spectroscopies that differ mainly by the number of photons and phonons involved in the scattering process \cite{long2002raman}. The first-order Raman process, in which only a single phonon is involved, is the dominant scattering process in samples with low defect concentrations. 

In a recent work, the first-order Raman spectra of 733 monolayer materials from the C2DB were calculated, and used as the basis for an automatic procedure for identifying a 2D material entirely from its experimental Raman spectrum\cite{taghizadeh2020library}. The Raman spectrum is calculated using third-order perturbation theory to obtain the rate of scattering processes involving creation/annihilation of one phonon and two photons, see Ref.~\cite{taghizadeh2020library} for details. The light field is written as $\boldsymbol{\mathcal{F}}(t) = \mathcal{F}_\mathrm{in}\mathbf{u}_\mathrm{in} \exp(-i\omega_\mathrm{in} t)+\mathcal{F}_\mathrm{out}\mathbf{u}_\mathrm{out} \exp(-i\omega_\mathrm{out} t)+$c.c., where $\mathcal{F}_\mathrm{in/out}$ and $\omega_\mathrm{in/out}$ denote the amplitudes and frequencies of the input/output electromagnetic fields, respectively. In addition, $\mathbf{u}_\mathrm{in/out}=\sum_i u_\mathrm{in/out}^i\mathbf{e}_i$ are the corresponding polarization vectors, where $\mathbf{e}_i$ denotes the unit vector along the $i$-direction with $i\in \lbrace x,y,z \rbrace$. Using this light field, the final expression for the Stokes Raman intensity involving scattering events by only one phonon reads \cite{taghizadeh2020library}
\begin{align}
\label{eq:ramanfinal}
&I(\omega) = I_0 \sum_\nu \frac{n_\nu+1}{\omega_\nu} \bigg|\sum_{ij} u_\mathrm{in}^i R_{ij}^\nu  u_\mathrm{out}^j \bigg|^2 \delta(\omega-\omega_\nu) \, .
\end{align}
Here, $I_0$ is an unimportant constant (since Raman spectra are always reported normalized), and $n_\nu$ is obtained from the Bose--Einstein distribution, i.e. $n_\nu \equiv (\exp[\hbar\omega_{\nu}/k_BT]-1)^{-1}$ at temperature $T$ for a Raman mode with energy $\hbar\omega_\nu$. Note that only phonons at the Brillouin zone center (with zero momentum) contribute to the one-phonon Raman processes due to momentum conservation. In Eq.~(\ref{eq:ramanfinal}), $R_{ij}^\nu$ is the Raman tensor for phonon mode $\nu$, which involves electron-phonon and dipole matrix elements as well as the electronic transition energies and the incident excitation frequency. Eq.~(\ref{eq:ramanfinal}) has been used to compute the Raman spectra of the 733 most stable, non-magnetic monolayers in C2DB for a range of excitation frequencies and polarization configurations. Note that the Raman shift $\hbar\omega$ is typically expressed in cm$^{-1}$ with 1 meV equivalent to 8.0655 cm$^{-1}$. In addition, for generating the Raman spectra, we have used a Gaussian [$G(\omega)=(\sigma\sqrt{2\pi})^{-1}\exp{(-\omega^2/2\sigma^2)}$] with a variance $\sigma=3$ cm$^{-1}$ to replace the Dirac delta function, which accounts for the inhomogeneous broadening of phonon modes. 

\begin{figure*}[t]
	\centering
	\includegraphics[width=0.9\textwidth]{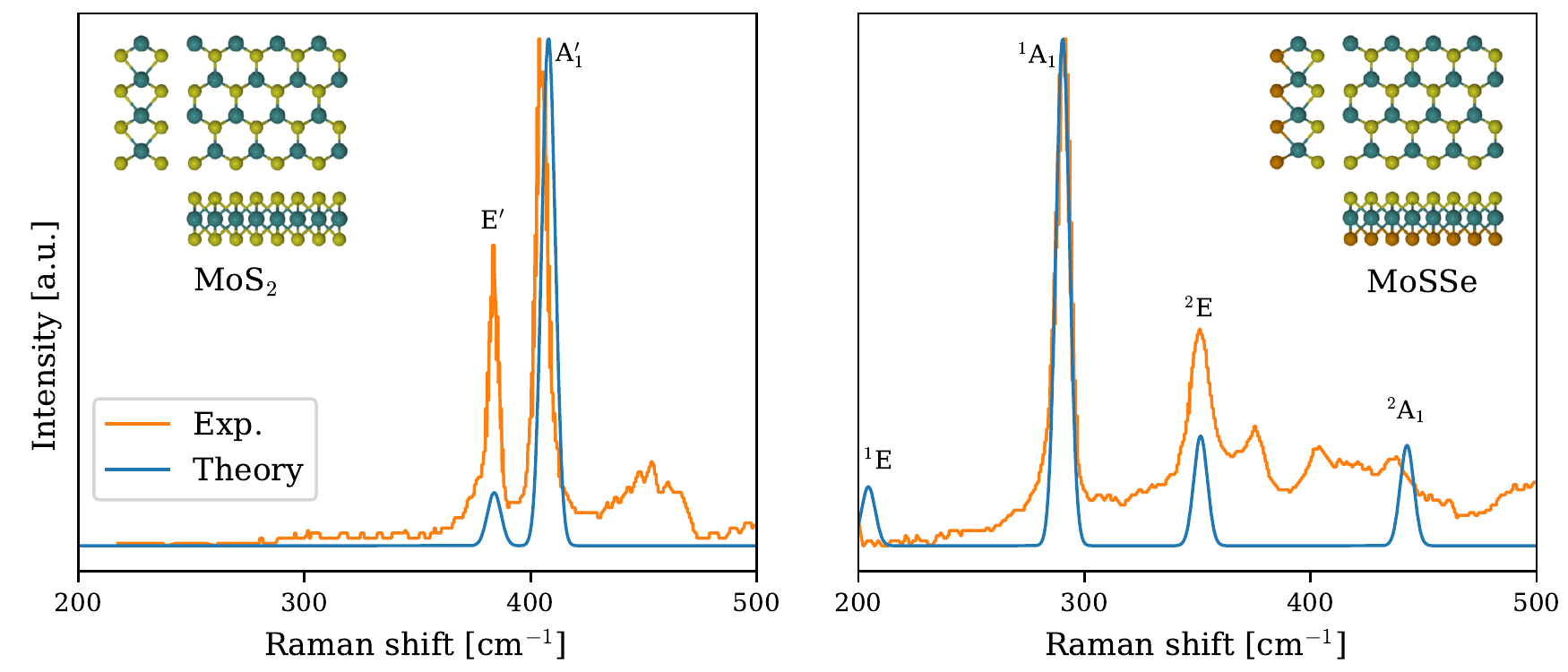}
	\caption[Raman spectrum]{Comparison of the calculated and experimental (extracted from Ref.~\cite{zhang2017janus}) Raman spectrum of MoS$_2$ (left) and MoSSe (right).  The excitation wavelength is 532 nm, and both the polarization of both the incoming and outgoing photons are along the $y$-direction. The Raman peaks are labeled according to the irreducible representations of the corresponding vibrational modes. Adapted from Ref.~\cite{taghizadeh2020library}.   } 
	\label{fig:raman_janus}
\end{figure*}
As an example, Fig.~\ref{fig:raman_janus} shows the calculated Raman spectrum of monolayer MoS$_2$ and the Janus monolayer MoSSe (see Sec. \ref{sec:janus}). Experimental Raman spectra extracted from Ref.~\cite{zhang2017janus} are shown for comparison. For both materials, good agreement between theory and experiment is observed for the peak positions and relative amplitudes of the main peaks. The small deviations can presumably be attributed to substrate interactions and defects in the experimental samples as well as the neglect of excitonic effects in the calculations. The qualitative differences between the Raman spectra can be explained by the different point groups of the materials ($C_{3v}$ and $D_{3h}$, respectively), see Ref.~\cite{taghizadeh2020library}. In particular, the lower symmetry of MoSSe results in a lower degeneracy of its vibrational modes leading to more peaks in the Raman spectrum.

\subsection{Second harmonics generation}

Nonlinear optical (NLO) phenomena such as harmonic generation, Kerr, and Pockels effects are of great technological importance for lasers, frequency converters, modulators, etc. In addition, NLO spectroscopy has been extensively employed to obtain insight into materials properties \cite{prylepa2018material} that are not accessible by e.g. linear optical spectroscopy. Among numerous nonlinear processes, second-harmonic generation (SHG) has been widely used for generating new frequencies in lasers as well as identifying crystal orientations and symmetries. 

Recently, the SHG spectrum was calculated for 375 non-magnetic, non-centrosymmetric semiconducting monolayers of the C2DB, and multiple 2D materials with giant optical nonlinearities were identified \cite{taghizadeh2020two}. In the SHG process, two incident photons at frequency $\omega$ generate an emitted photon at frequency of $2\omega$. Assume that a mono-harmonic electric field written $\boldsymbol{\mathcal{F}}(t) = \sum_{i} \mathcal{F}_i \mathbf{e}_i e^{-i\omega t}$+c.c. is incident on the material, where $\mathbf{e}_i$ denotes the unit vector along direction $i\in\{x,y,z\}$. The electric field induces a SHG polarization density $\mathbf{P}^{(2)}$, which can be obtained from the quadratic susceptibility tensor $\chi_{ijk}^{(2)}$, 
\begin{align}
P_i^{(2)}(t) = \epsilon_0 \sum_{jk} \chi_{ijk}^{(2)}(\omega,\omega) \mathcal{F}_i \mathcal{F}_j e^{-2i\omega t} + \textrm{c.c.} \, ,
\end{align}
where $\epsilon_0$ denotes the vacuum permittivity. $\chi_{ijk}^{(2)}$ is a symmetric (due to intrinsic permutation symmetry i.e. $\chi_{ijk}^{(2)}=\chi_{ijk}^{(2)}$) rank-3 tensor with at most 18 independent elements. Furthermore, similar to the piezoelectric tensor, the point group symmetry reduces the number of independent tensor elements. 

In the C2DB, the quadratic susceptibility is calculated using density matrices and perturbation theory \cite{aversa1995nonlinear, taghizadeh2017linear} with the involved transition dipole matrix elements and band energies obtained from DFT. The use of DFT single-particle orbitals implies that excitonic effects are not accounted for. The number of empty bands included in the sum over bands was set to three times the number of occupied bands. The width of the Fermi-Dirac occupation factor was set to $k_BT=50$ meV, and a line-shape broadening of $\eta = 50$ meV was used in all spectra. Furthermore, time-reversal symmetry was imposed in order to reduce the $\mathbf{k}$-integrals to half the BZ. For various 2D crystal classes, it was verified by explicit calculation that the quadratic tensor elements fulfill the expected symmetries, e.g. that they all vanish identically for centrosymmetric crystals.

\begin{figure*}[t]
	\centering
	\includegraphics[width=0.9\textwidth]{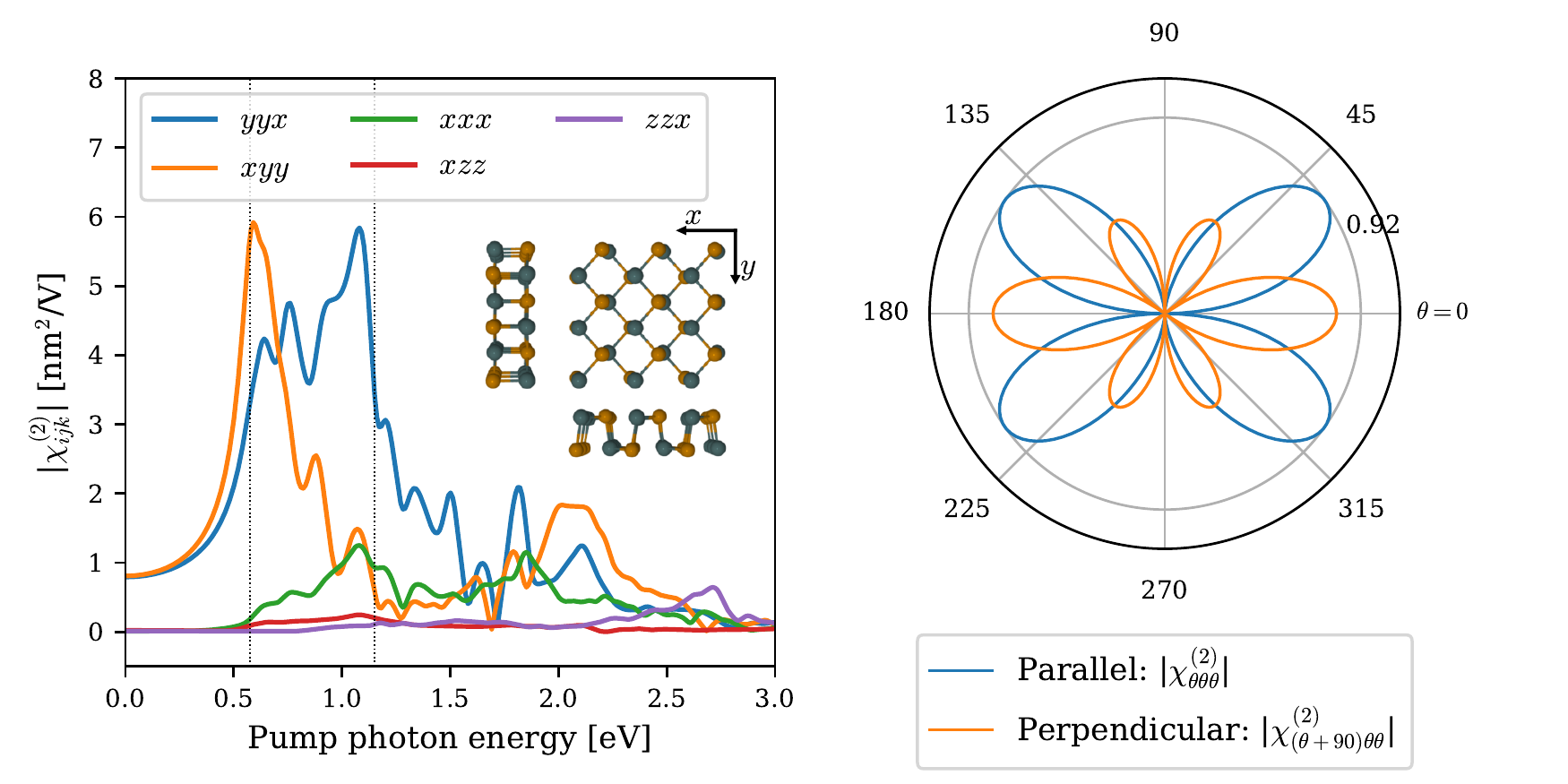}
	\caption[SHG spectrum]{(Left panel) SHG spectra of monolayer Ge$_2$Se$_2$, where only non-vanishing independent tensor elements are shown. The vertical dashed lines mark $\hbar\omega=E_g/2$ and $\hbar\omega=E_g$, respectively. The crystal structure of Ge$_2$Se$_2$ structure is shown in the inset. (Right panel) The rotational anisotropy of the static ($\omega=0$) SHG signal for parallel (blue) and perpendicular (red) polarization configurations with $\theta$ defined with respect to the crystal $x$-axis.} 
	\label{fig:shg_spectra}
\end{figure*}

As an example, the calculated SHG spectra for monolayer Ge$_2$Se$_2$ is shown in Fig.~\ref{fig:shg_spectra} (left panel). Monolayer Ge$_2$Se$_2$ has 5 independent tensor elements, $\chi_{xxx}^{(2)}$, $\chi_{xyy}^{(2)}$, $\chi_{xzz}^{(2)}$, $\chi_{yyx}^{(2)}=\chi_{yxy}^{(2)}$, and $\chi_{zzx}^{(2)}=\chi_{zxz}^{(2)}$, since it is a group-IV dichalcogenide with an orthorhombic crystal structure (space group 31 and point group $C_{2v}$). Note that, similar to the linear susceptibility, the bulk quadratic susceptibility (with SI units of m/V) is ill-defined for 2D materials (since the volume is ambiguous) \cite{taghizadeh2020two}. Instead, the unambiguous \textit{sheet} quadratic susceptibility (with SI units of m$^2$/V) is evaluated. In addition to the frequency-dependent SHG spectrum, the angular dependence of the static ($\omega=0$) SHG intensity at normal incidence for parallel and perpendicular polarizations (relative to the incident electric field) is calculated, see Fig.~\ref{fig:shg_spectra} (right panel). Such angular resolved SHG spectroscopy has been widely used for determining the crystal orientation of 2D materials. The calculated SHG spectra for all non-vanishing inequivalent polarization configurations and their angular dependence, are available in the C2DB.

\begin{figure}[t]
	\centering
	\includegraphics[width=0.45\textwidth]{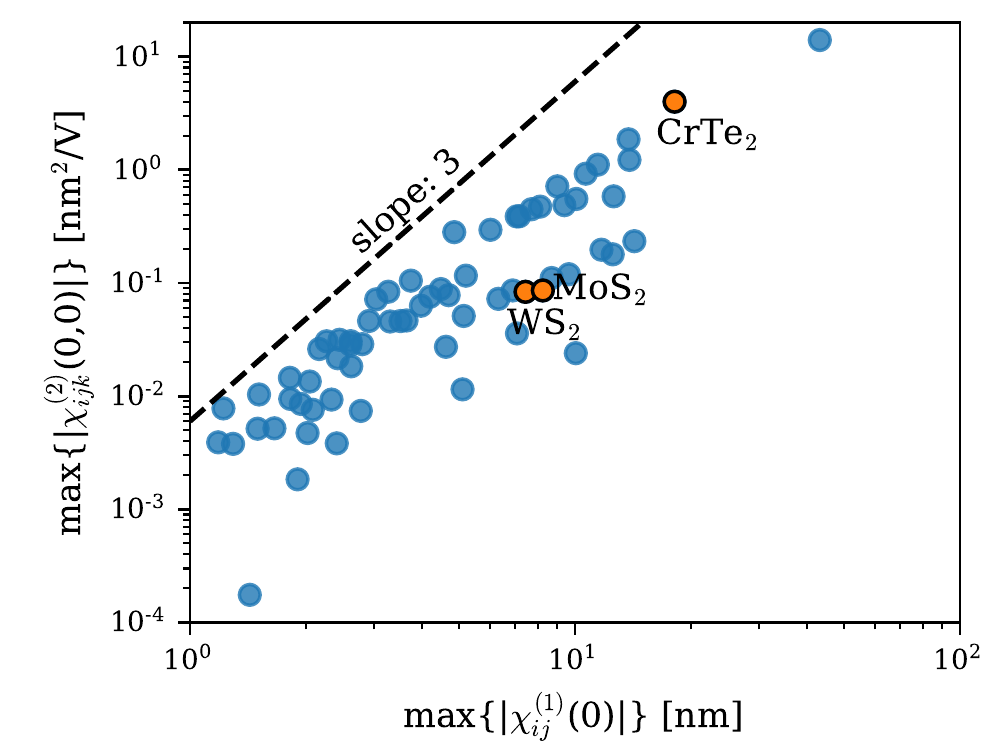}
	\caption[SHG]{Scatter plot (double log scale) of the static sheet quadratic susceptibility $|\chi_{ijk}^{(2)}|$ versus the static sheet linear susceptibility $|\chi_{ij}^{(1)}|$ for 67 TMDCs (with chemical formula MX$_2$ and space group 187). A few well known materials are highlighted.} 
	\label{fig:shg_linear}
\end{figure}

Since C2DB has already gathered various material properties of numerous 2D materials, it provides a unique opportunity to investigate interrelations between different material properties. For example, the strong dependence of the quadratic optical response on the electronic band gap was demonstrated on basis of the C2DB data \cite{taghizadeh2020two}. As another example of a useful correlation, the static quadratic susceptibility is plotted versus the static linear susceptibility for 67 TMDCs (with formula MX$_2$, space group 187) in Fig.~\ref{fig:shg_linear}. Note that for materials with several independent tensor elements, only the largest is shown. There is a very clear correlation between the two quantities. This is not unexpected as both the linear and quadratic optical responses are functions of the transition dipole moments and transition energies. More interestingly, the strength of the quadratic response seems to a very good approximation to be given by a universal constant times the linear susceptibility to the power of three (ignoring polarisation indices), i.e.
\begin{align}
\chi^{(2)}(0,0)\approx A \chi^{(1)}(0)^3,
\end{align}
where $A$ is only weakly material dependent. Note that this scaling law is also known in classical optics as semi-empirical Miller's rule for non-resonant quadratic responses \cite{miller1964optical}, which states that the second order electric susceptibility is proportional to the product of the first-order susceptibilities at the three frequencies involved.

%Similar analyses can be performed for other material classes or properties to extract further trends and correlations from our data set. 

\section{Machine learning properties}\label{sec:ML}
In recent years, material scientists have shown great interest in exploiting the use of machine learning (ML) techniques for predicting materials properties and guiding the search for new materials. ML is the scientific study of algorithms and statistical models that computer systems can use to perform a specific task without using explicit instructions but instead relying on patterns and inference. Within the domain of materials science, one of the most frequent problems is the mapping from atomic configuration to material property, which can be used e.g. to screen large material spaces in search of optimal candidates for specific applications. \cite{Schmidt2019,Zhuo2018}

In the ML literature, the mathematical representation of the input observations is often referred to as a fingerprint. Any fingerprint must satisfy a number of general requirements. \cite{Faber2015} In particular, a fingerprint must be
\begin{description}
\item{\textit{Complete:}} The fingerprint should incorporate all the relevant input for the underlying problem, i.e. materials with different properties should have different fingerprints.
\item{\textit{Compact:}} The fingerprint should contain no or a minimal number of features redundant to the underlying problem. This includes being invariant to rotations, translations and other transformations that leave the properties of the system invariant.
\item{\textit{Descriptive:}} Materials with similar target values should have similar fingerprints.
\item{\textit{Simple:}} The fingerprint should be efficient to evaluate. In the present context, this means that calculating the fingerprint should be significantly faster than calculating the target property.
\end{description}

\begin{figure*}[t]
	\centering
	\includegraphics[width=\linewidth]{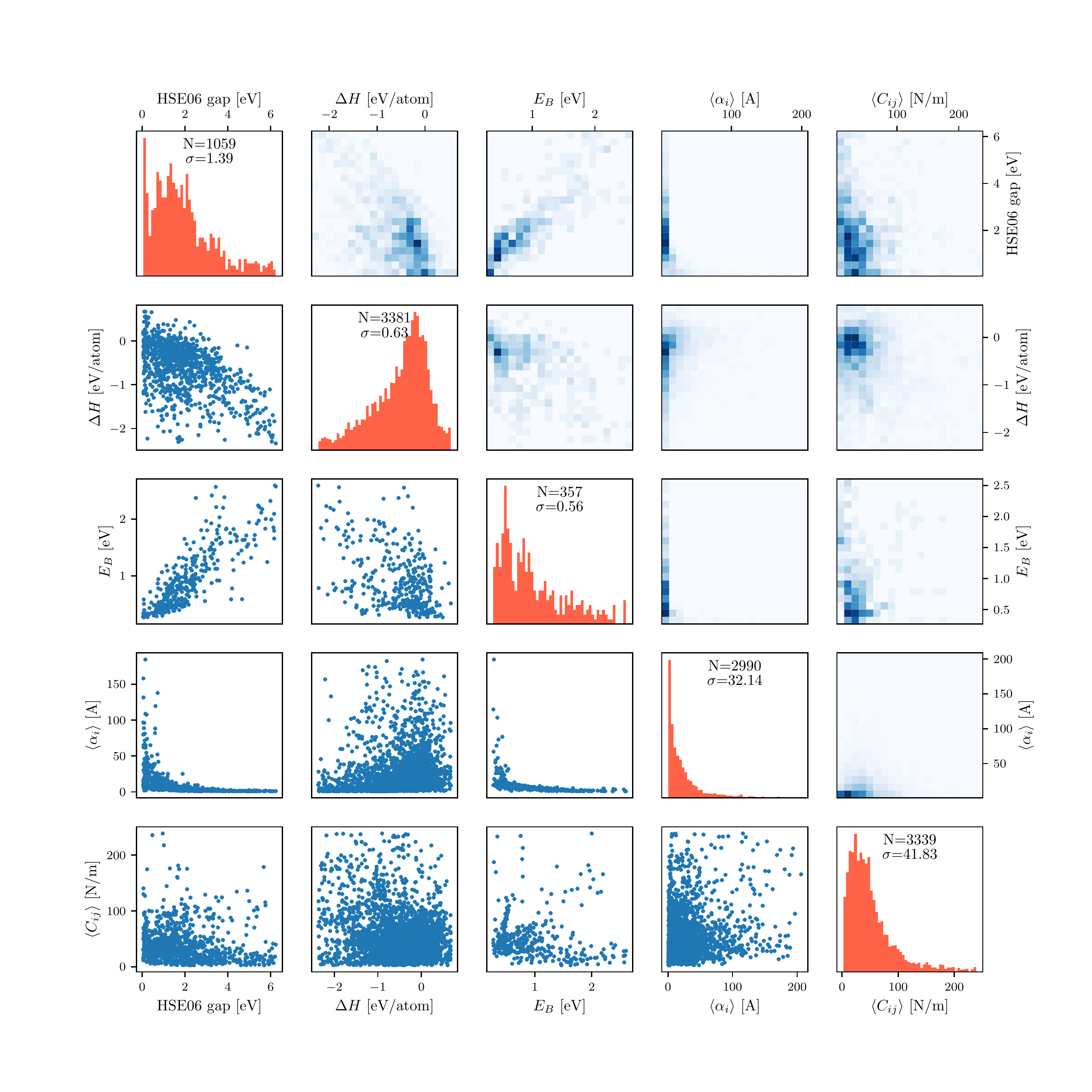}
	\caption{Pair-plot of selected properties from C2DB. The diagonal contains the single property histograms. Below the diagonal are two-property scatter plots showing the correlation between properties and above the diagonal are two-property histograms. properties include the HSE06 band gap, the PBE heat of formation ($\Delta H$), the exciton binding energy ($E_B$) calculated from the Bethe-Salpeter equation (BSE), the in-plane static polarisability calculated in the random phase approximation (RPA) and averaged over the $x$ and $y$ polarisation directions ($\langle \alpha_i\rangle $), and the in-plane Voigt modulus ($\langle C_{ii}\rangle $) defined as $\frac{1}{4} \left( C_{11}+C_{22}+2C_{12} \right)$, where $C_{ij}$ is a component of the elastic stiffness tensor.} 
	\label{fig:ML_pairplot}
\end{figure*}

Several types of atomic-level materials fingerprints have been proposed in the literature, including general purpose fingerprints based on atomistic properties\cite{ward2016general,ghiringhelli2015big} possibly encoding information about the atomic structure, i.e. atomic positions\cite{rupp2012fast,Faber2015,huo2018unified}, and specialized fingerprints tailored for specific applications (materials/properties)\cite{jorgensen2018machine,rajan2018machine}. 

The aim of this section is to demonstrate how the C2DB may be utilized for ML-based prediction of general materials properties. Moreover, the study serves to illustrate the important role of the fingerprint for such problems. The 2D materials are represented using three different fingerprints: two popular structural fingerprints and a more advanced fingerprint that encodes information about the the electronic structure via the projected density of states (PDOS). The target properties include the HSE06 band gap, the PBE heat of formation ($\Delta H$), the exciton binding energy ($E_B$) obtained from the many-body Bethe-Salpeter equation (BSE), the in-plane static polarisability calculated in the random phase approximation (RPA) averaged over the $x$ and $y$ polarisation directions ($\langle \alpha_i\rangle $), and the in-plane Voigt modulus ($\langle C_{ii}\rangle $) defined as $\frac{1}{4} \left( C_{11}+C_{22}+2C_{12} \right)$, where $C_{ij}$ is a component of the elastic stiffness tensor in Mandel notation.

To introduce the data, Figure \ref{fig:ML_pairplot} shows pair-plots of the dual-property relations of these properties. The plots in the diagonal show the single-property histograms, whereas the off-diagonals show dual-property scatter plots below the diagonal and histograms above the diagonal. Clearly, there are only weak correlations between most of the properties, with the largest degree of correlation observed between the HSE06 gap and exciton binding energy. The lack of strong correlations motivates the use of machine learning for predicting the properties. 

The prediction models are build using the Ewald sum matrix and many-body tensor representation (MBTR) as structural fingerprints. The Ewald fingerprint is a version of the simple Coulomb matrix fingerprint\cite{rupp2012fast} modified to periodic systems \cite{Faber2015}. The MBTR encodes first, second and third order terms like atomic numbers, distances and angles between atoms in the system \cite{huo2018unified}. As an alternative to the structural fingerprints, a representation based on the PBE projected density of states (PDOS) is also tested. This fingerprint (to be published) encodes the coupling between the PDOS at different atomic orbitals and the distance between atoms. Since this fingerprint requires a DFT-PBE calculations to be performed, additional features derivable from the DFT calculation can be added to the fingerprint. In this study, the PDOS fingerprint is amended by the PBE band gap. The latter can in principle be extracted from the PDOS, but its explicit inclusion improves performance (see below). 

A Gaussian process regression using a simple Gaussian kernel with a noise component is used as learning algorithm. The models are trained using 5-fold cross validation on a training set consisting of $80\%$ of the materials with the remaining $20\%$ held aside as test data. Prior to training the model, the input space is reduced to 50 features using principal component analysis (PCA). This step is neccesary to reduce the huge number of features in the MBTR fingerprint to a manageable size. Although this is not required for the Ewald and PDOS fingerprints, we perform the same feature reduction in all cases. The optimal number of features depends on the choice of fingerprint, target property and learning algorithm, but for consistency 50 PCA components are used for all fingerprints and properties in this study. 

\begin{figure}[t]
	\centering
	\includegraphics[width=\linewidth]{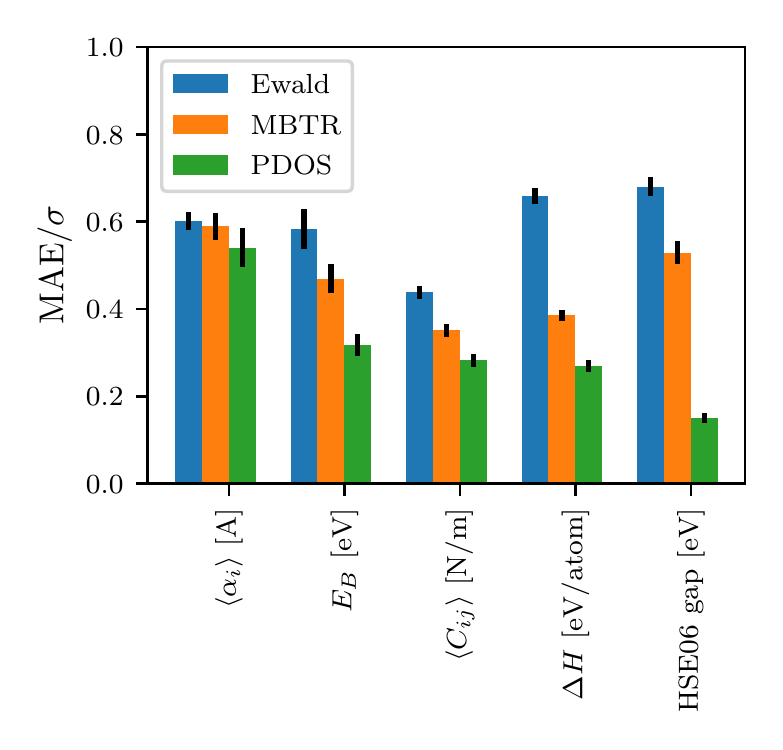}
	\caption{Prediction scores (MAE normalized to standard deviation of property values) for the test sets of selected properties using a Gaussian process regression.} 
	\label{fig:ML_scores}
\end{figure}

Figure \ref{fig:ML_scores} shows the prediction scores obtained for the 5 properties using the three different fingerprints. The employed prediction score is the mean absolute error of the test set normalized by the standard deviation of the property values (standard deviations are annotated in the diagonal plots in Fig.~\ref{fig:ML_pairplot}). In general, the PDOS fingerprint outperforms the structural fingerprints. The difference between prediction scores is smallest for the static polarisability $\langle \alpha_i\rangle $ and largest for the HSE06 gap. It should be stressed that although the evaluation of the PBE-PDOS fingerprint is significantly more time consuming than the evaluation of the structural fingerprints, it is still much faster than the evaluation of all the target properties. Moreover, structural fingerprints require the atomic structure, which in turns requires a DFT structure optimization (unless the structure is available by other means).  

\begin{figure*}[t]
	\centering
	\includegraphics[width=\linewidth]{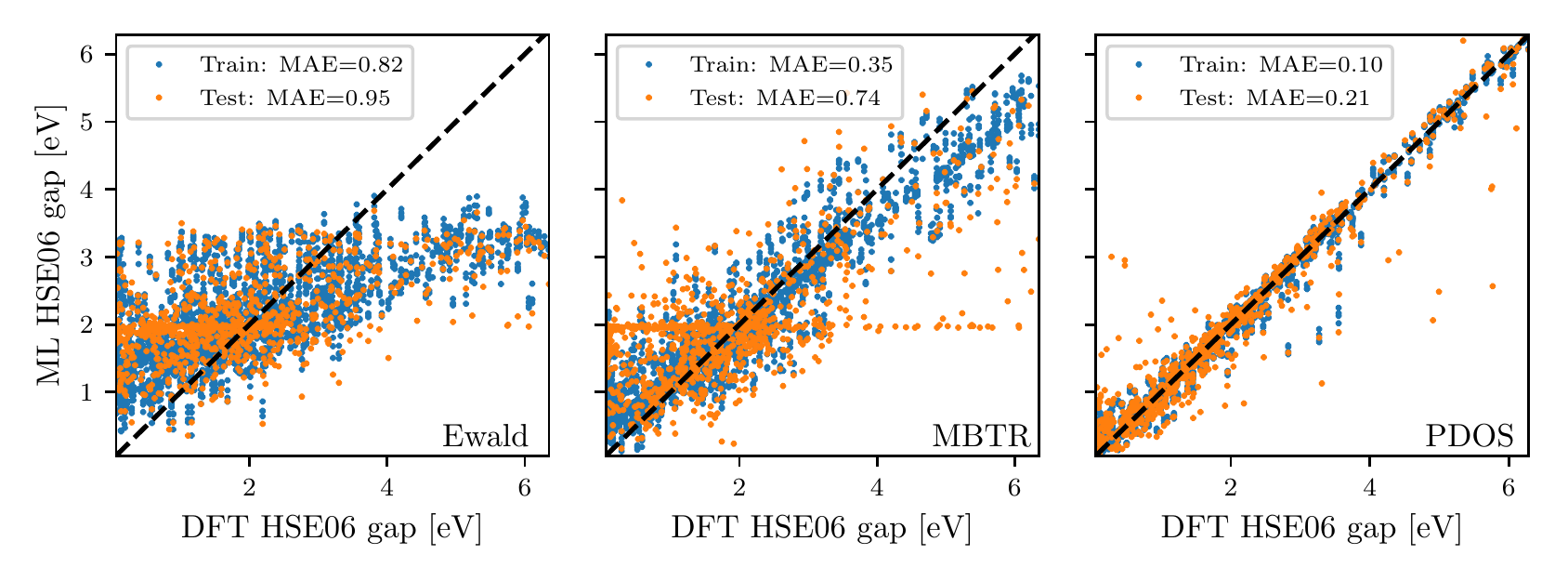}
	\caption{ML predicted HSE06 gap values vs. true values for Ewald, MBTR and PDOS fingerprints with MAE's for train and test set included. The PDOS is found to perform significantly better for the prediction of HSE06 gap.}
	\label{fig:ML_scatter}
\end{figure*}

The HSE06 band gap shows the largest sensitivity to the employed fingerprint. To elaborate on the HSE06 results, Fig. \ref{fig:ML_scatter} shows the band gap predicted using each of the three different fingerprints plotted against the true band gap. The mean absolute errors on the test set is 0.95 eV and 0.74 eV for Ewald and MBTR fingerprints, respectively, while the PDOS significantly outperforms the other fingerprints with a test MAE of only 0.21 eV. This improvement in prediction accuracy is partly due to the presence of the PBE gap in the PDOS fingerprint. However, our analysis shows that the pure PDOS fingerprint without the PBE gap still outperforms the structural fingerprints. Using only the PBE gap as feature results in a test MAE of 0.28 eV. 

The current results show that the precision of ML-based predictions are highly dependent on the type of target property and the chosen material representation. For some properties, the mapping between atomic structure and property is easier to learn while others might require more/deeper information, e.g. in terms of electronic structure fingerprints. Our results clearly demonstrate the potential of encoding electronic structure information into the material fingerprint, and we anticipate more work on this relevant and exciting topic in the future.  

\section{Summary and outlook}

We have documented a number of extensions and improvements of the Computational 2D Materials Database (C2DB) made in the period 2018-2020. The new developments include: (i) A refined and more stringent workflow for filtering prospect 2D materials and classifying them according to their crystal structure, magnetic state and stability. (ii) Improvements of the methodology used to compute certain challenging properties such as the full stiffness tensor, effective masses, G$_0$W$_0$ band structures, and optical absorption spectra. (iii) New materials including 216 MXY Janus monolayers and 574 monolayers exfoliated from experimentally known bulk crystals. In addition, ongoing efforts to systematically obtain and characterize bilayers in all possible stacking configurations as well as point defects in the semiconducting monolayers, have been described. (iv) New properties including exfoliation energies, spontaneous polarisations, Bader charges, piezoelectric tensors, infrared (IR) polarisabilities, topological invariants, magnetic exchange couplings, Raman spectra, and second harmonic generation spectra. It should be stressed that the C2DB will continue to grow as new structures and properties are being added, and thus the present paper should not be seen as a final report on the C2DB but rather a snapshot of its current state.

In addition to the above mentioned improvements relating to data quantity and quality, the C2DB has been endowed with a comprehensive documentation layer. In particular, all data presented on the C2DB website are now accompanied by an information field that explains the meaning and representation (if applicable) of the data and details how it was calculated thus making the data easier to understand, reproduce, and deploy.

The C2DB has been produced using the Atomic Simulation Recipes (ASR) in combination with the GPAW electronic structure code and the MyQueue task and workflow scheduling system. The ASR is a newly developed Python-based framework designed for high-throughput materials computations. The highly flexible and modular nature of the ASR and its strong coupling to the well established community-driven ASE project, makes it a versatile framework for both high- and low-throughput materials simulation projects. The ASR and the C2DB-ASR workflow are distributed as open source code. A detailed documentation of the ASR will be published elsewhere.     

While the C2DB itself is solely concerned with the properties of perfect monolayer crystals, ongoing efforts focus on the systematic characterisation of homo-bilayer structures as well as point defects in monolayers. The data resulting from these and other similar projects will be published as separate, independent databases, but will be directly interlinked with the C2DB making it possible to switch between them in a completely seamless fashion. These developments will significantly broaden the scope and usability of the C2DB+ (+ stands for associated databases) that will help theoreticians and experimentalists to navigate one of the most vibrant and rapidly expanding research fields at the crossroads of condensed matter physics, photonics, nanotechnology, and chemistry.

\section{Acknowledgments}
The Center for Nanostructured Graphene (CNG) is sponsored by the Danish National Research Foundation, Project DNRF103. This project has received funding from the European Research Council (ERC) under the European Union’s Horizon 2020 research and innovation program grant agreement No 773122 (LIMA). T.D. acknowledges financial support from the German Research Foundation (DFG Projects No. DE 2749/2-1).

\section*{References}
\bibliographystyle{iopart-num}
\bibliography{references}

\end{document}